\documentclass[11pt]{article}
\usepackage[symbol]{footmisc}

\usepackage[left=2.5cm,top=2.5cm,right=2cm,bottom=2.5cm]{geometry}
\usepackage{amsmath, amssymb}

\usepackage{graphicx,subfigure,epsfig,epsf,color}
\usepackage{slashed}
\usepackage{feynmf}
\usepackage{mathrsfs}
\usepackage{graphicx}
\usepackage{graphics}
\usepackage{epstopdf}
\usepackage{subfigure}
\usepackage{amsfonts}
\usepackage{float}
\usepackage{sectsty}
\usepackage{hyperref}
\usepackage{cite}
\usepackage{pdflscape}
\usepackage{lipsum}
\usepackage{graphicx}
\usepackage{amsthm,amsmath,amssymb}
\usepackage{mathrsfs}
\usepackage{subfigure}
\usepackage{float}
\usepackage{lscape}
\usepackage{array}
\usepackage{mathrsfs}
\usepackage{booktabs}
\usepackage{multicol}
\usepackage{amsthm,amsmath,amssymb}
\usepackage{mathrsfs}
\usepackage{subfigure}
\usepackage{float}
\usepackage{lscape}
\usepackage{array}
\usepackage{mathrsfs}
\usepackage{booktabs}
\usepackage{multicol}
\usepackage[figuresright]{rotating}
\begin{document}
	\date{}

\title{Electromagnetic counterparts of high-frequency gravitational waves in a rotating laboratory frame system and their detection}
	\maketitle
	\begin{center}
		Fang-Yu Li$~^{a,}$\footnote{fangyuli@cqu.edu.cn, corresponding author},~Hao Yu~$^{a,}$\footnote{yuhaocd@cqu.edu.cn, corresponding author},~Jin Li~$^{a,}$\footnote{cqujinli1983@cqu.edu.cn},~Lian-Fu Wei~$^{b,}$\footnote{lfwei@swjtu.edu.cn},~Qing-Quan Jiang~$^{c,}$\footnote{qqjiang@cwnu.edu.cn}
	\end{center}
	
	\begin{center}
		$^a$ Physics Department, Chongqing University, Chongqing 401331, China\\
		$^b$ Information Quantum Technology Laboratory, International Cooperation Research Center of China
		Communication and Sensor Networks for Modern Transportation, School of Information Science and Technology,
		Southwest Jiaotong University, Chengdu 610031, China\\
		$^c$ Institute of Theoretical Physics,
		China West Normal University,
		Nanchong, 637009, China
	\end{center}
	\vspace*{0.25in}
	\begin{abstract}
		We consider the perturbative photon flows (PPFs, i.e., electromagnetic (EM) counterparts) generated by the EM resonance response to high-frequency gravitational waves (HFGWs) with additional polarization states in a rotating laboratory frame system. It is found that when the propagating direction of HFGWs and the symmetrical axis of the laboratory frame system are the same, the PPFs have the maximum value. In this case, using the rotation (the rotation of the azimuth $\phi$) of the EM detection system, all six possible polarization states of HFGWs can be separated and displayed. For the current experimental conditions, it is quite prospective to detect the PPFs generated by the HFGWs predicted in the braneworld models, the primordial black hole theories, the interaction mechanism between astrophysical plasma and intense EM radiation, etc., due to the large amplitudes (or high spectral densities) and spectral characteristics of these HFGWs. Detecting the primordial HFGWs from inflation faces great challenges at present, but it is not impossible.	
	\end{abstract}
	
	\section{Introduction}

The interaction cross section between gravitational waves (gravitons) and matter (elementary particles) is extremely weak, which makes all matter almost transparent to GWs. Therefore, GWs can pass through huge distances and carry oldest and furthest information in cosmology and astrophysics. Since the direct detection of GWs is very difficult, the successful detection of GWs in intermediate frequency ($\nu\sim 30$ Hz to $450$ Hz) is undoubtedly great achievements (see Ref.~\cite{LIGOScientific:2016aoc} and relevant subsequent references given by the LIGO Scientific Collaboration and the Virgo Collaboration), which opens up a new epoch of GW astronomy and multi-messenger astronomy.
On the other hand, such GWs are located in an interesting but special frequency range, and the corresponding duration of the signal in the detector is very short. Thus, how to detect continuous GWs and GWs in other frequency bands is of great significance and an urgent affair at present. In fact, many mainstream cosmological models, high-energy astrophysical processes, and even high-energy physical experiments have predicated the existence of the HFGWs in the GHz band or even higher frequencies, and the corresponding amplitudes would be expected to be $h\sim$ $10^{-23}$ to $10^{-39}$~\cite{Aggarwal:2020olq,Franciolini:2022htd,Saito:2021sgq,Wu:2008hw}. {These GWs are expected to exhibit a broad frequency spectrum, spanning from extremely low frequencies ($\nu\sim10^{-17}$ Hz or below) to high frequencies ($\nu\sim10^9$ Hz or higher). Notably, the high-frequency components of these GWs are believed to carry crucial information about cosmology and high-energy astrophysics. The observations of GWs across different frequency bands can complement one another in unveiling various aspects of the universe. The observations of GWs at higher frequencies have the potential to reveal new physics beyond the Standard Model of particle physics, as highlighted in Ref.~\cite{Aggarwal:2020olq}. Furthermore, these observations could provide insights into exotic objects within the universe, such as primordial black holes and boson stars, as well as cosmological events that occurred during the early stages of the universe, including phase transitions, preheating after inflation, quantum oscillons, cosmic strings, thermal fluctuations after reheating, and other phenomena.

Moreover, the frequencies of the GWs (KK-gravitons) expected by braneworld scenarios~\cite{Clarkson:2006pq,Seahra:2004fg}, and the GWs predicted by the interaction between astrophysical plasma and intense electromagnetic waves~\cite{Servin:2003cf}, are anticipated to range from $\nu\sim10^9$ Hz to $\nu\sim10^{12}$ Hz or higher. The amplitudes (strains) of these HFGWs would be expected to be in the range of $h\sim10^{-21}$ to $h\sim10^{-27}$. In some studies, it is even predicted that the frequencies of HFGWs originating from the coherent oscillation of electron-positron pairs and fields~\cite{Han:2013ska} or magnetars~\cite{Wen:2016swd} could exceed $10^{19}$ Hz. The detection of HFGWs within these frequency ranges would provide valuable insights into the nature of the universe, including the properties of extra dimensions, the interaction mechanism between astrophysical plasma and electromagnetic fields, and the potential existence of exotic sources such as the coherent oscillations of electron-positron pairs and fields or magnetars.}

Since the frequencies of these GWs are far beyond the detection range of traditional ground-band GW detectors, their observation and detection face new challenges and opportunities. {For example, the main challenge in observing the HFGWs predicted by inflation models lies in their inherent weakness, characterized by extremely small amplitudes (strains) and low spectral densities. For conventional inflation models, the expected amplitudes would be on the order of $h\sim10^{-30}$ or even smaller. If the amplitudes of the aforementioned HFGWs are less than $h\sim10^{-31}$, the current detection schemes would be unable to detect them unless there are significant advancements in detection principles and experimental technology. Moreover, in addition to this challenge, there are other obstacles to detecting the HFGWs predicted by inflation models, including interference from noise, cosmic variance, and theoretical uncertainties. These factors make it difficult to differentiate between the HFGWs expected by inflation models and other sources. Nevertheless, despite these challenges, the potential insights into the early universe and fundamental physics make the pursuit of HFGW detection an exceptionally exciting area of research. Therefore, in order to detect HFGWs with the current experimental technology, we need to propose new principles and schemes, e.g. see Refs.~\cite{Arvanitaki:2012cn,Franciolini:2022htd,Domcke:2020yzq,Aggarwal:2020olq,Vermeulen:2020djm,Galliou1,Goryachev1,Aggarwal:2020umq,Ejlli:2019bqj,Liu:2023mll,Zheng:2022yzo,Li:2017jcz}.} By the way, in Ref.~\cite{Ejlli:2019bqj} the authors studied the electromagnetic detection of the HFGWs with frequencies from $10^{14}$ Hz to $10^{18}$ Hz. This frequency band is almost the upper limit of the current HFGW detection. In this scheme, large-scale background static magnetic field is an obvious advantage, because it can generate significant spatial accumulation effects of the EM signal. Moreover, each signal photon generated by the electromagnetic response to the ultra-high frequency band will have larger energy. In this case, the requirements for suppressing the thermal noise and the signal accumulation time of the signal photon flow can be greatly relaxed. Recently, Liu et al. have proposed that the nearby planets can be used to detect HFGWs~\cite{Liu:2023mll}. In the scheme, they obtained the first limits from the existing low-Earth-orbit satellite for specific frequency bands and projected the sensitivities for the future more-dedicated detections. There are also some other important principles and schemes for detecting HFGWs, which have been discussed and reviewed in detail in Refs.~\cite{Aggarwal:2020olq} and~\cite{Franciolini:2022htd}. For more details, one can refer to these references and so we will not fully cover these researches here.

In Ref.~\cite{Li:2017jcz}, we investigated the electromagnet (EM) response to the HFGWs having additional polarization states. In general, these HFGWs might have, at most, six polarization states: $\oplus$-type and $\otimes$-type polarizations (the tensor-mode gravitons); $x$-type and $y$-type polarizations (the vector-mode gravitons); $b$-type and $e$-type polarizations (the scalar-mode gravitons). We assume that the HFGWs propagate along the $z$-direction, and the $\oplus$-type and $\otimes$-type polarizations among the six polarization states satisfy the transverse-traceless (TT) gauge condition in general relativity (GR). The frame system in which the $\oplus$-type and $\otimes$-type polarizations of GWs satisfy the TT gauge condition can be called the standard GW frame system. In such an ideal frame system, we have proved that it is possible to distinguish and probe the tensor-mode, vector-mode and scalar-mode gravitons by the EM response to HFGWs~\cite{Li:2017jcz}.

However, any GW detection systems fixed on the Earth have a rotation period of $24$ hours to the GWs sources having determinate local spacial location due to the rotation of the Earth. Therefore, an important scientific problem would be what directly observable quantities of the EM resonance response to HFGWs in the laboratory frame system are. Moreover, we need to figure out how to distinguish and display the EM perturbations produced by different polarization states of HFGWs in the laboratory frame system. In this paper, we shall give a general relationship between the EM perturbations and the rotation of the laboratory frame system, including the corresponding physical effects in different conditions.

The paper is organized as follows. In sec.~\uppercase\expandafter{\romannumeral2}, we show general relationship among the tensor polarizations (the $\oplus$-type and $\otimes$-type polarizations) and the non-tensor polarizations (the $x$-type, $y$-type, $b$-type, and $l$-type polarizations) of HFGWs. We also introduce the rotation into our three-dimensional synchro-resonance system (3DSR system). In sec.~\uppercase\expandafter{\romannumeral3}, we  briefly introduce the principle and scheme of the 3DSR system and calculate the EM perturbations (the perturbative photon flows (PPFs), i.e., the signal photon flows) generated by the EM resonance response to HFGWs in the 3DSR system. In sec.~\uppercase\expandafter{\romannumeral4}, we give the signal and noise analysis with the help of the numerical estimation, and study the distinguishability and detectability of the PPFs in the $3$DSR system. Our conclusions are summarized in sec.\uppercase\expandafter{\romannumeral5}, including some discussions for future prospects.\\
	\section{Perturbative EM fields generated by different polarization states of HFGWs}
	The ``monochromatic components" of GWs which have six polarization states and propagate along the $z$-direction in the standard GW frame system can be written as
	\begin{equation}
		\begin{split}\label{1}
			h_{ij}=\left({A_{\oplus}e_{ij}^{\oplus}+A_{\otimes}e_{ij}^{\otimes}+A_xe_{ij}^x+A_ye_{ij}^y+A_be_{ij}^b}{+A_le_{ij}^l}\right) e^{i\left(k_gz-\omega_gt\right)},
		\end{split}	
	\end{equation}
	where $k_g$ and $\omega_g$ are wave number and angular frequency of the GWs, respectively. Each polarization mode can be defined as the following matrix form~\cite{Nishizawa:2009bf}
	\begin{align}\label{2}
		e_{ij}^{\bigoplus} &= \left( \begin{array}{ccc}
			1 & 0 & 0 \\
			0 & -1 & 0 \\
			0 & 0 & 0 \end{array} \right)
		&e_{ij}^{\bigotimes} =& \left( \begin{array}{ccc}
			0 & 1 & 0 \\
			1 & 0 & 0 \\
			0 & 0 & 0 \end{array} \right)\notag\\
		e_{ij}^x &= \left( \begin{array}{ccc}
			0 & 0 & 1 \\
			0 & 0 & 0 \\
			1 & 0 & 0 \end{array} \right)
		&e_{ij}^y =& \left( \begin{array}{ccc}
			0 & 0 & 0 \\
			0 & 0 & 1 \\
			0 & 1 & 0 \end{array} \right)\notag\\
		e_{ij}^b &= \left( \begin{array}{ccc}
			1 & 0 & 0 \\
			0 & 1 & 0 \\
			0 & 0 & 0 \end{array} \right)
		&e_{ij}^l =& \sqrt{2} \left( \begin{array}{ccc}
			0 & 0 & 0 \\
			0 & 0 & 0 \\
			0 & 0 & 1 \end{array} \right).
	\end{align}
	The labels $\oplus$, $\otimes$, $x$, $y$, $b$ and $l$ represent $\oplus$-type, $\otimes$-type polarizations (tensor-mode gravitons), $x$-type, $y$-type polarizations (vector-mode gravitons), and $b$-type, $l$-type polarizations (scalar-mode gravitons), respectively.
	
Since the weak GWs can be regarded as a small perturbation $h_{\mu\nu}$ to the background spacetime $\eta_{\mu\nu}$, from Eqs.~(\ref{1}) and (\ref{2}), the metric tensor of the weak GWs can be given as	
	\begin{align}\label{newb3}
		g_{\mu\nu} &=\eta_{\mu\nu}+h_{\mu\nu}= \left( \begin{array}{cccc}
			-1 & 0 & 0 & 0 \\
			0 & 1+h_{\oplus}+h_b & h_{\otimes} & h_x\\
			0 & h_{\otimes} & 1-h_{\oplus}+h_b & h_y\\
			0 & h_x & h_y & 1+h_l\end{array} \right).
	\end{align}

\begin{figure}[H]
		\centering
		\renewcommand{\figurename}{Fig.}
		\includegraphics[width=21cm]{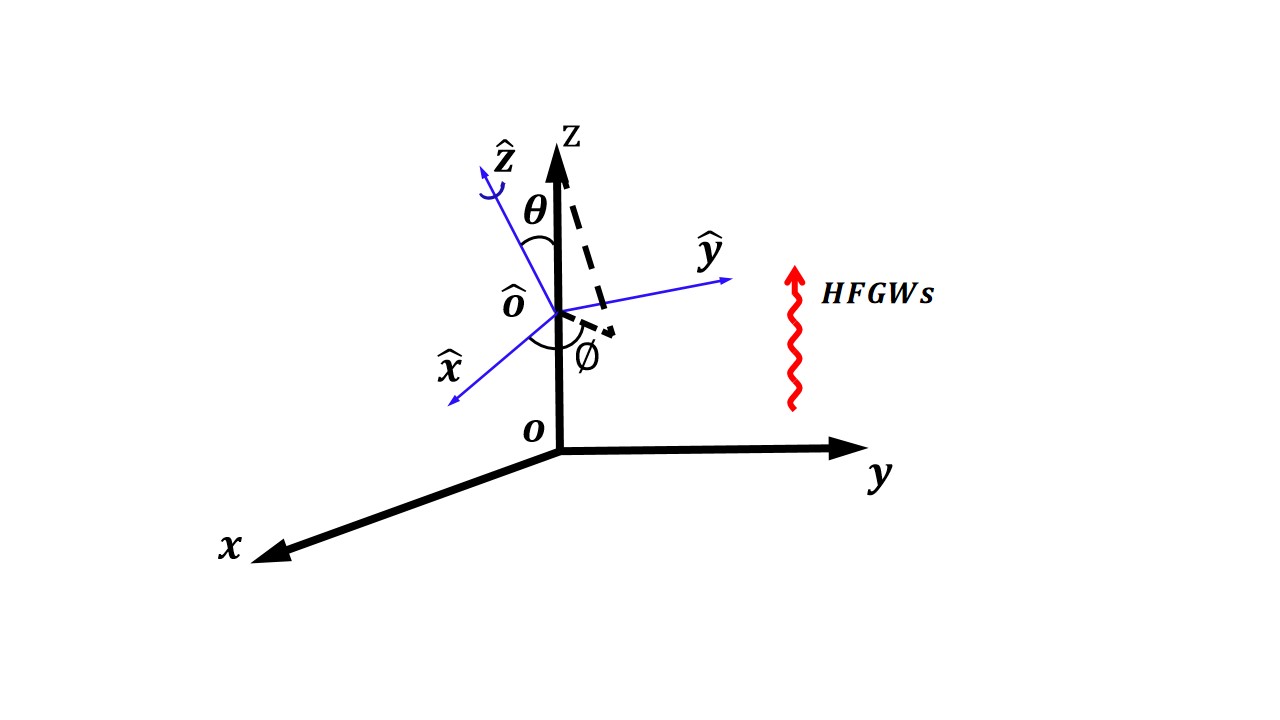}
		\caption{The standard GW frame system $(x,y,z)$ (for HFGWs) and the laboratory frame system $(\hat{x},\hat{y},\hat{z})$ (for the 3DSR system). The HFGWs propagate along the $z$-direction in the standard GW frame system. Suppose that the laboratory frame system for the detector (i.e., the 3DSR system) is rotated by angles $(\theta,\phi)$.}	\label{11111}
	\end{figure}

Note that Eqs.~(\ref{1}), (\ref{2}) and (\ref{newb3}) are the component forms in the standard GW frame system. However, since any detection systems fixed on the Earth have a rotation period of $24$ hours to the coherent GW sources with determinate local spacial location, the detectable and observable physical quantities are not usually consistent with the forms in the standard GW frame system. In this work, we study the three-dimensional synchro-resonance system (3DSR system) (see Refs.~\cite{Li:2017jcz,Woods2011,Wen:2014hia,Tong:2008rz,Li:2008qr}) in the laboratory frame system, so the symmetrical axis of the 3DSR system may deviate from the propagating direction of HFGWs (see Fig.~\ref{11111}). Moreover, in our 3DSR system, the observer should be at rest in the static magnetic fields. Thus, the observable quantities should be the projects of the physical quantities as a tensor on tetrads of the observer's world line.
	
According to Fig.~\ref{11111}, the general relation between the standard GW frame system $(x,y,z)$ and the laboratory frame system $(\hat{x},\hat{y},\hat{z})$ can always be written as the following simple form
	\begin{equation}\label{3}
		\begin{split}
			\hat{x} &= x\cos \theta\cos \phi+y\cos \theta \sin \phi-z\sin \theta, \\
			\hat{y} &= -x\sin \phi+y\cos \phi, \\
			\hat{z} &= x\sin \theta\cos \phi+y\sin \theta \sin \phi+z\cos \theta.
		\end{split}
	\end{equation}
However, seeking the general relation between the perturbative EM fields generated by HFGWs in the standard GW frame system $(x,y,z)$ and the ones in the laboratory frame system $(\hat{x},\hat{y},\hat{z})$ is complicated. Fortunately, our 3DSR system has a special and determined direction, namely, the positive direction of the symmetrical axis of the Gaussian beam (GB) (the $\hat{z}$-direction in the laboratory frame system, see Fig.~\ref{11111}). It is  shown~\cite{Li:2008qr,Tong:2008rz} that only when the propagating direction (the $z$-direction in the standard GW frame system) of the HFGWs and the $\hat z$-direction are the same (i.e., $\theta=0$ in Fig.~\ref{11111}), the PPFs generated by HFGWs have maximum value. In this case, the transformation between the PPFs in the standard GW frame system and the ones in the laboratory frame system will only depend on the azimuth angle $\phi$ and so it will be greatly simplified (see below). If they are perpendicular or opposite to each other, then the PPFs will be one or two orders of magnitude smaller than the maximum value. Thus, we can determine the propagating direction of HFGWs by observing the peak moments of the PPFs.
	
Moreover, because the 3DSR system is a small-scale EM detection system (whose volume is on the order of cubic meters), there is no any difficulty with the rotation of the 3DSR system in principle. Based on the laser gyro positioning and some other direction positioning and tracking methods, the spatial orientation of the 3DSR system can be effectively adjusted to the best or near optimal detection direction for the possible HFGW signals.
	
According to Eqs.~(\ref{1}) to (\ref{3}), when the propagating direction of HFGWs and the $\hat z$-direction coincide (i.e., $\theta=0$ in Fig.~\ref{11111}), the polarization states $h_{ij}$ in the standard GW frame system $(x,y,z)$ and the ones in the laboratory frame system $(\hat{x},\hat{y},\hat{z})$ have following relation:
	\begin{equation}\label{4}
		\begin{split}
			{h}_{11} &= h_\oplus\cos 2\phi +h_\otimes \sin 2\phi +h_b,\\
			{h}_{12} &= {h}_{21} = -h_\oplus\sin 2\phi +h_\otimes\cos 2\phi ,\\
			{h}_{22} &= -h_\oplus\cos 2\phi -h_\otimes\sin 2\phi +h_b,\\
			{h}_{13} &= {h}_{31} = h_x \cos \phi +h_y\sin \phi ,\\
			{h}_{23} &= {h}_{32} = - h_x\sin \phi+h_y\cos \phi ,\\
			{h}_{33} &= h_l.	
		\end{split}
	\end{equation}
Base on Eq.~(\ref{4}), one can find that:
	
$(1)$ Once $\phi=0$ (the standard GW frame system and the laboratory frame system are perfectly coincident), then Eq.~(\ref{4}) is reduced to Eqs.~(\ref{1}), (\ref{2}) and (\ref{newb3})~\cite{Li:2017jcz}.
	
$(2)$ There is only the conversion between the tensor-mode gravitons (the $\oplus$-type and $\otimes$-type polarizations), and the conversion between the vector-mode gravitons (the $x$-type and $y$-type polarizations). But, there is no conversion between tensor-mode gravitons and non-tensor-mode gravitons. And there is no conversion between the vector-mode gravitons (the $x$-type and $y$-type polarizations) and the scalar-mode gravitons (the $b$-type and $l$-type polarizations). {The tensor mode, vector mode, and scalar mode correspond to gravitons with spin 2, spin 1, and spin 0, respectively. Since spin is an intrinsic property of particles, the rotation of coordinates alone does not induce changes in particle spin. Therefore, these results are consistent with our understanding of particle spin.}
	
$(3)$ The $b$-type and $l$-type polarizations (scalar-mode gravitons) are only located at the $xx$-, $yy$-, and $zz$-components of the metric tensor of HFGWs, respectively. They are independent on the azimuth $\phi$ when the coordinates are rotated. These properties will be very useful and important for probing the $b$-type and $l$-type polarizations of HFGWs (see below).
	
According to Eqs.~(\ref{1}), (\ref{2}) and (\ref{newb3}), using the electrodynamics equations in curved spacetime:
	\begin{equation}\label{5}
		\frac{\partial}{\partial x^{\nu}}\left[\sqrt{-g} g^{\mu \alpha} g^{\nu\beta}\left(\hat F_{\alpha \beta}^{(0)}+\tilde{F}_{\alpha \beta}^{(1)}\right)\right]=0,
	\end{equation}	
	\begin{equation}\label{6}	
		\nabla_{\mu} F_{\nu \alpha}+\nabla_{\nu} F_{\alpha \mu}+\nabla_{\alpha} F_{\mu \nu}=0,   	
	\end{equation}
we have obtained the first-order perturbative EM fields $\tilde{F} _{\alpha,\beta}^{(1)}$ generated by HFGWs in the standard GW frame system~\cite{Li:2017jcz}. Here, $\hat{F}_{\alpha,\beta}^{(0)}$ is the static background EM field tensor, ``~0~'' indicates the background EM fields, ``~1~'' indicates the first-order perturbation, ``$~\hat~~$'' denotes the static EM fields, and ``$~\tilde ~~$" represents the time-dependent perturbative EM fields.
	
Combining with Eqs.~(\ref{1}) to (\ref{6}) and Ref.~\cite{Li:2017jcz}, the perturbative EM fields (the EM counterparts) generated by the EM response to HFGWs, in the {general frame system} (in MKS units) can be given by
	\begin{align}\label{7}
		\tilde{E}_{x}^{(1)} =& -\frac{i}{2} k_{g} c z \Big[ \left(-h_\oplus\sin 2\phi +h_\otimes\cos 2\phi  \right) \hat{B}_{x}^{(0)}                              \notag\\
		&-\left(h_\oplus\cos 2\phi +h_\otimes\sin 2\phi +\frac{1}{2} h_l\right) \hat{B}_{y}^{(0)}                   \notag \\
		&+\left(-h_x\sin \phi +h_y\cos \phi \right)\hat{B}_{z}^{(0)}\Big],\\\label{8}
		\tilde{B}_{y}^{(1)} =& -\frac{i}{2} k_{g} z \Big[ \left(-h_\oplus\sin 2\phi +h_\otimes\cos 2\phi  \right) \hat{B}_{x}^{(0)}                             \notag\\
		&-\left(h_\oplus\cos 2\phi +h_\otimes\sin 2\phi +\frac{1}{2} h_l\right) \hat{B}_{y}^{(0)}                  \notag \\
		&+\left(-h_x\sin \phi +h_y\cos \phi \right)\hat{B}_{z}^{(0)}\Big],            \\\label{9}
		\tilde{E}_{y}^{(1)} =&\frac{i}{2} k_{g} cz\Big[\left(h_{b}-h_{\oplus}\cos 2\phi -h_\otimes\sin 2\phi +h_{l}\right) \hat{B}_{x}^{(0)}      \notag \\
		& +\left(-h_\oplus\sin 2\phi +h_\otimes\cos 2 \phi\right) \hat{B}_{y}^{(0)}                              \notag \\
		& +\left(h_x\cos \phi +h_y\sin \phi \right) \hat{B}_{z}^{(0)}\Big],                         \\\label{10}
		\tilde{B}_{x}^{(1)} =&-\frac{i}{2} k_{g} z\Big[\left(h_{b}-h_{\oplus}\cos 2\phi -h_\otimes\sin 2\phi +h_{l}\right) \hat{B}_{x}^{(0)}      \notag \\
		& +\left(-h_\oplus\sin 2\phi +h_\otimes\cos 2\phi \right) \hat{B}_{y}^{(0)}                              \notag \\
		& +\left(h_x\cos \phi +h_y\sin \phi \right) \hat{B}_{z}^{(0)}\Big],                           \\\label{11}
		\tilde{E}_{z}^{(1)}=&\tilde{B}_{z}^{(1)}=0.
	\end{align}
If $\phi=0$, Eqs.~(\ref{7}) to (\ref{10}) are reduced to the perturbative EM fields to the background static magnetic fields in the standard GW frame system. Therefore, the expressions for the perturbative EM fields and the corresponding PPFs in the standard GW frame system are the special forms of the EM response to HFGWs. Here, we have neglected the perturbative EM fields in the opposite direction of the propagation of HFGWs, because they are very weak (they do not have the space-accumulation effect) or absent~\cite{Boccaletti1970}. Moreover, here we only consider the EM response of the static magnetic fields to HFGWs and do not study the case of the static electric fields. This is because from the perspective of the experiment, it is more realistic to produce a strong static magnetic field (such as the 10-Tesla magnetic field) than to generate an equivalent static electric field.
	
However, to detect the power flows of the EM signal (i.e., signal photon flows or PPFs) produced by HFGWs in the laboratory frame system, we face two problems. On the one hand, the PPFs represented by Eqs.~(\ref{7}) to (\ref{10}) are actually the second-order perturbative EM effects produced by HFGWs, because their strengths are proportional to the square of the amplitude ($h\sim10^{-23}$ to $10^{-30}$ or less) of HFGWs. Since the PPFs are very faint, it is difficult to probe them directly under the current laboratory conditions. On the other hand, the rotation of the Earth will lead to some uncertainty of the PPFs in the laboratory frame system. Fortunately, such two problems, in principle, can be solved.

\begin{figure}[H]
		\centering
		\renewcommand{\figurename}{Fig.}
		\includegraphics[width=12cm]{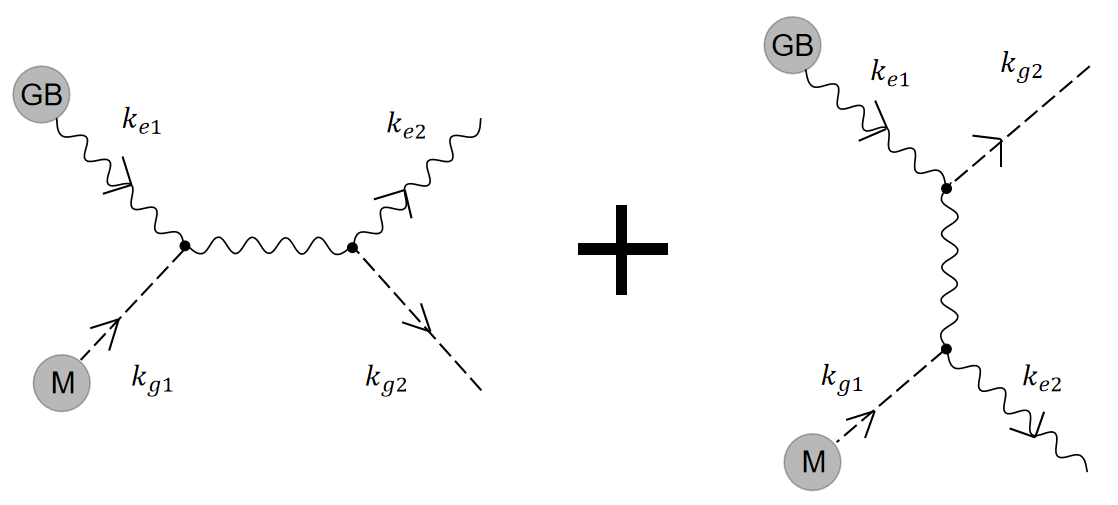}
		\caption{Feynman diagrams of the resonant interaction between the gravitons (HFGWs) $K_{g1}$ and the photons (GB) $K_{e1}$ in the static magnetic field $M$ (virtual photon background acting as a ``catalyst''). Here, ``resonant'' means that they have the same frequency. Then, the interaction cross section is much larger than the one of the direct interaction between the gravitons (HFGWs) and the virtual photon background $M$. The latter interaction without the GB is often called (pure) inverse Gertsenshtein effect~\cite{Gertsenshtein1962,ZeldovichA,ZeldovichB}.}\label{fig2}
	\end{figure}

For the first problem, we can introduce a Gaussian-type photon flow [i.e., Gaussian beam (GB)] into the background static magnetic fields. Then, utilizing the EM resonance response to HFGWs, we can obtain first-order PPFs~\cite{Li:2017jcz,Li:2008qr,Woods2011} [e.g., see the following Eqs.~(\ref{23}) and (\ref{24})]. Since the first-order PPFs are proportional to the amplitude $h$ of HFGWs rather than the square of the amplitude, the first-order PPFs will be much larger than the second-order PPFs and then the difficulty of detecting PPFs is greatly reduced. The quantum picture of this process can be expressed by Feynman diagrams (see Fig.~\ref{fig2}), which describe the resonant interaction between the photons (i.e., GB) and the gravitons (HFGWs) in a background (static magnetic field) of virtual photons. The static magnetic field as a catalyst can greatly increase the interaction cross section between the photons and the gravitons. Thus, introducing a GB into the background EM field can effectively compensate to the weakness of the HFGW amplitudes when we use the EM resonance response to HFGWs to detect HFGWs.

It should be pointed out that the introduction of the GB will cause large background noise photon flows (BPFs). However, since the first-older PPFs and
the BPFs have very different physical behaviors in special local areas of the 3DSR system,
such as intensity distribution, propagating direction, decay rate, wave impedance,
etc. (see below), the 3DSR system has a very low standard quantum limit~\cite{Stephenson:2009zz} so that it is always possible to distinguish and display the PPFs and the BPFs in the 3DSR system.
	
{For the second problem, according to the general relation between the perturbative EM fields in the laboratory frame system and the ones in the standard GW frame system, it is possible to determine the observable quantities of the perturbative EM fields, including the PPFs generated by HFGWs in the laboratory frame system.}

\section{First-order perturbative photon flows (PPFs) in the laboratory frame system}

\subsection{3DSR system}

The 3DSR system consists of background static magnetic fields, a Gaussian-type photon flow (i.e., GB), a fractal membrane (an equivalent microwave lens~\cite{ShahzadAnwar,Wen2112,zhou1112,Li2009}), and a weak photon flow detector. Here, ``three-dimensional'' means that under the resonance condition ($\omega_e = \omega_g$), not only the longitudinal first-order PPFs along the propagating direction (i.e., the $z$-direction) of HFGWs can be produced, but the transverse first-order PPFs (propagating along the $x$- and $y$-directions) can also be generated (see below). Based on the very low standard quantum limit (high-sensitivity) of the 3DSR system and the different physical behaviors between the transverse first-order PPFs and the BPFs, distinguishing and probing all possible six polarization states of HFGWs would be, in principle, possible.
	
In general, GBs can have different forms, such as circular polarization of fundamental frequency modes or elliptical polarization of higher-order modes. Without loss of generality, here we consider a GB which has non-vanishing longitudinal magnetic component $\tilde B_z^{(0)}$ and the magnetic component $\tilde B_z^{(0)}$ has the standard form of the fundamental frequency circular mode in the framework of quantum electronics~\cite{Yariv1989}.
	
Combining with the Helmholtz equation, the Gauss's law for the free EM field (without sources) ($\nabla \cdot \tilde E^{(0)}=\nabla \cdot \tilde B^{(0)}=0$) and $\tilde E^{(0)}=\frac{ic}{k_e}\nabla \times \tilde B^{(0)} $, the EM components of the GB in the 3DSR system have the following forms:
	\begin{align}
		\tilde B_x^{(0)}&=\psi_{bx}=0\label{new12},\\
		\tilde B_y^{(0)}&=\psi_{by}=-\int\frac{\partial \psi_{bz}}{\partial z} dy\label{new13},\\
		\tilde B_z^{(0)}&=\psi_{bz}=\psi=\frac{\psi_0}{\sqrt{1+(z/f)^2}}\text{exp}\left(\frac{-r^2}{W^2}\right)
		\notag\\
		&\times\exp \left\{i\left[\left(k_{e} z-\omega_{e} t\right)-\tan ^{-1} \frac{z}{f}\right.\right.
		\left.\left.+\frac{k_{e} r^{2}}{2 R}+\delta\right]\right\}\label{new14},\\
		\tilde E_x^{(0)}&=\psi_{ex}=\frac{ic}{k_e}\left(\frac{\partial \psi_{bz}}{\partial y}-\frac{\partial \psi_{by}}{\partial z}\right)
		\label{new15},\\
		\tilde E_y^{(0)}&=\psi_{ey}=-\frac{ic}{k_e}\frac{\partial \psi_{bz}}{\partial x}
		\label{new16},\\
		\tilde E_z^{(0)}&=\psi_{ez}=-\frac{ic}{k_e}\frac{\partial \psi_{by}}{\partial x}
		\label{new17},
	\end{align}
where $\psi_0$ is the amplitude of the longitudinal magnetic field of the GB. The other parameters are given as $f=\pi \frac{\rm W_0^2}{\lambda_e}$, $ W=W_0$$\sqrt{1+(\frac{z}{f})^2}$, and $R=z+\frac{f^2}{z}$. Here, $W_0$ is the minimum spot radius, $R$ is the curvature radius of wave front of the GB at $z$, $\omega_e$ is the angular frequency, $\lambda_e$ is the EM wavelength, the $z$-axis is the symmetrical axis of the GB, and $\delta$ is a phase factor. 

\begin{figure}[H]
	\centering
	\renewcommand{\figurename}{Fig.}
	\includegraphics[width=17.5cm]{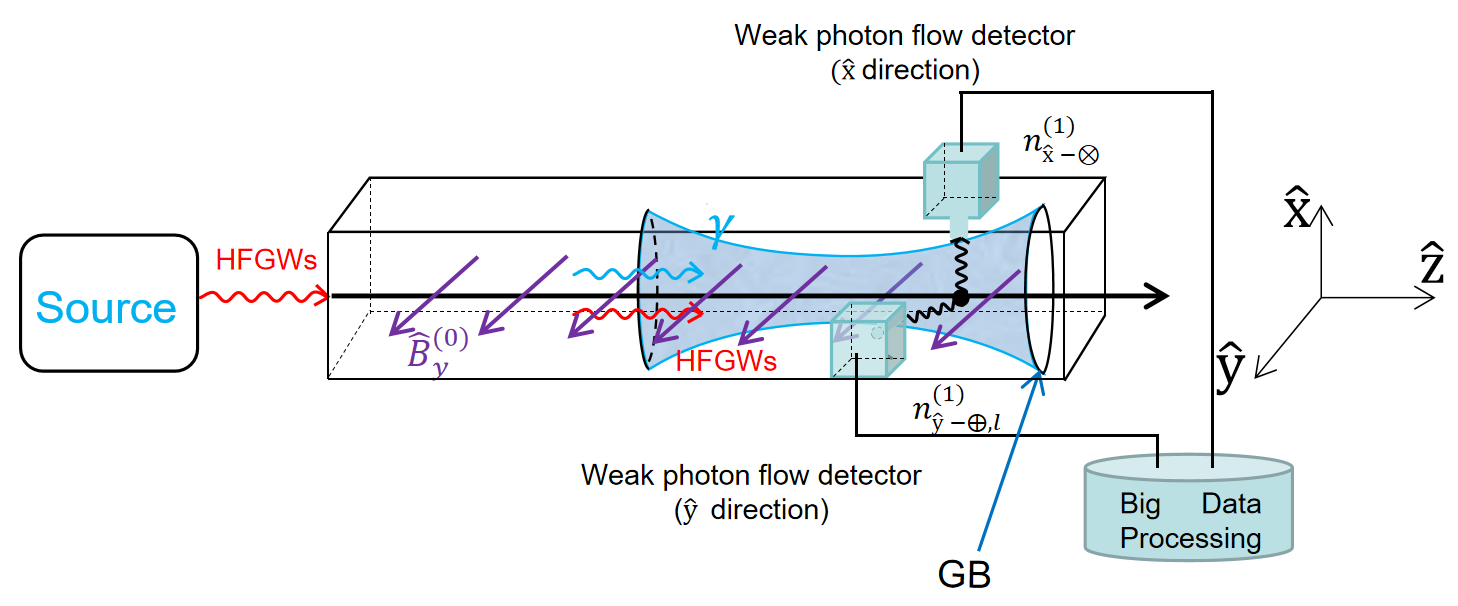}
	\caption{3DSR system with the coupling between the transverse static magnetic field $\hat{B}_y^{(0)}$ and the GB. Here, $n_{\hat x-\otimes}^{(1)}$ [see Eq.~(\ref{23})] is the PPF propagating along the $\hat x$-direction, while $n_{\hat y-\oplus,l}^{(1)}$  [see Eq.~(\ref{24})] is the PPF propagating along the $\hat y$-direction. The PPFs are generated by the pure $\otimes$-type polarization and the combined state of the $\oplus$-type and $l$-type polarizations of HFGWs, respectively.}\label{F3}	
\end{figure}

\begin{figure}[H]
	\centering
	\renewcommand{\figurename}{Fig.}
	\includegraphics[width=17.5cm]{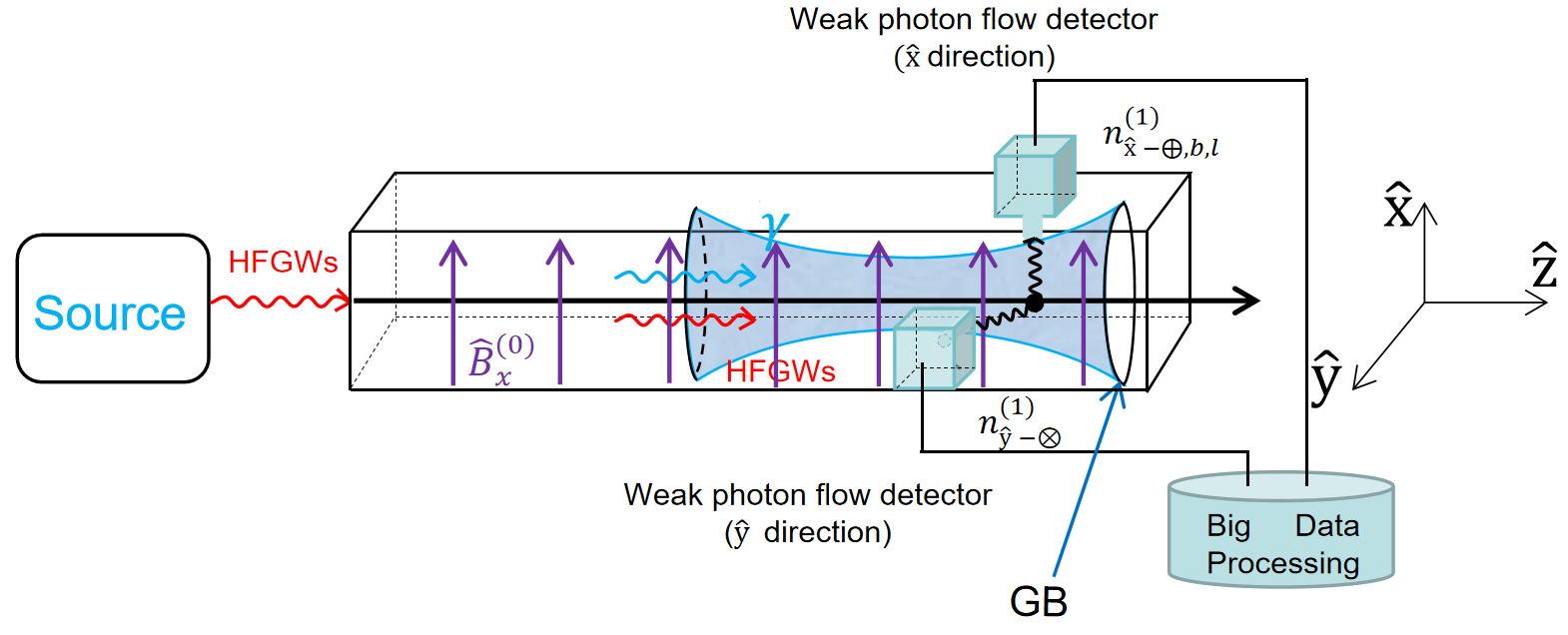}
	\caption{3DSR system with the coupling between the transverse static magnetic field $\hat{B}_x^{(0)}$ and the GB. Here, $n_{\hat x-\oplus,b,l}^{(1)}$  [see Eq.~(\ref{29})] is the PPF propagating along the $\hat{x}$-direction, while $n_{\hat y-\otimes}^{(1)}$  [see Eq.~(\ref{30})] is the PPF propagating along the $\hat{y}$-direction. The PPFs are generated by the combined state of the $\oplus$-type, $b$-type and $l$-type polarizations and the pure $\otimes$-type polarization of HFGWs, respectively.} \label{F4}			
\end{figure}

\begin{figure}[H]
	\centering
	\renewcommand{\figurename}{Fig.}
	\includegraphics[width=17.5cm]{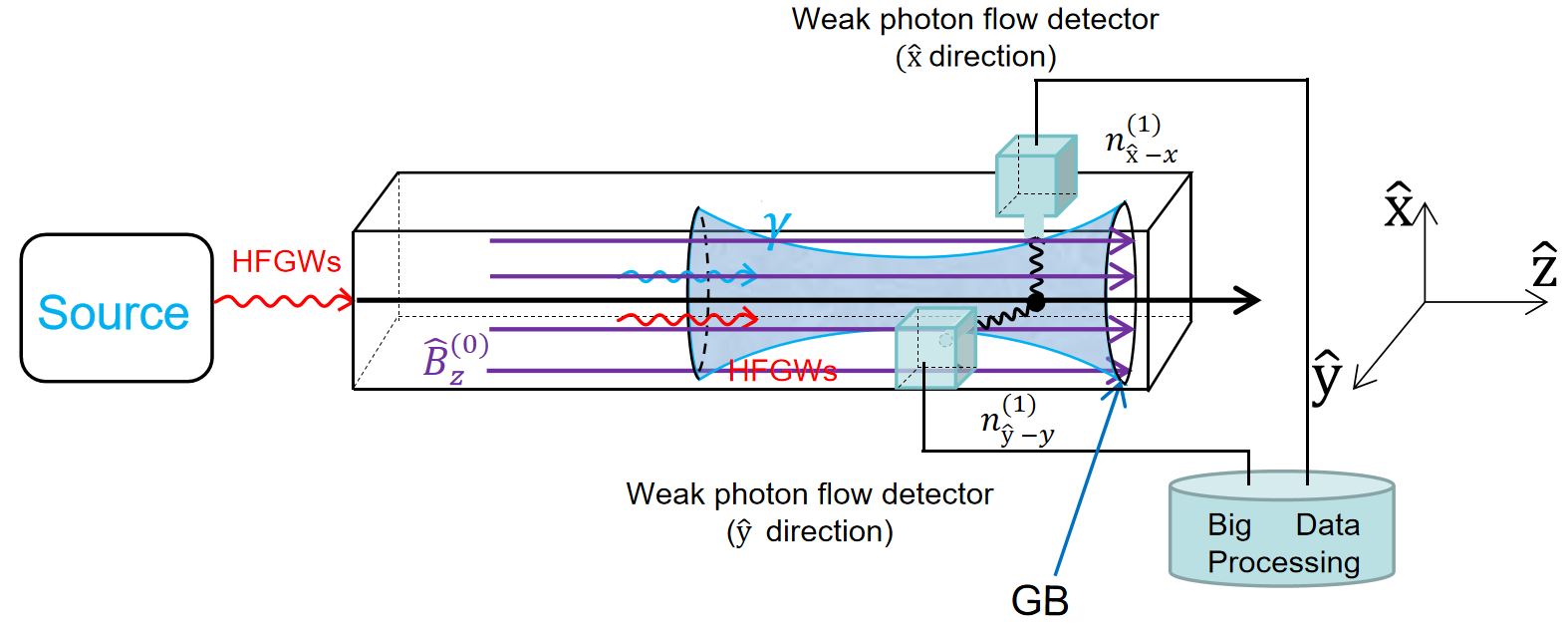}
	\caption{3DSR system with the coupling between the longitudinal static magnetic field $\hat{B}_z^{(0)}$ and the GB, Here, $n_{\hat x-x}^{(1)}$  [see Eq.~(\ref{33})] is the PPF propagating along the $\hat x$-direction, while $n_{\hat y-y}^{(1)}$ [see Eq.~(\ref{34})] is the PPF propagating along the $\hat y$-direction. The PPFs are generated by the pure $x$-type polarization and the pure $y$-type polarization of HFGWs, respectively.} 	\label{F5}		
\end{figure}

In the 3DSR system, the static magnetic field $\hat{B}^{(0)}(\hat{B}_x^{(0)},\hat{B}_y^{(0)},\hat{B}_z^{(0)})$ can point in an arbitrary direction and the direction can be adjusted (see Figs.~\ref{F3}, \ref{F4}, and \ref{F5}). It is worth noting that even if the 3DSR system is positioned in the optimal detection direction (the propagating direction (the $z$-direction) of HFGWs in the 3DSR system is along the $\hat z$-axis) by the laser gyro positioning and some other direction positioning and tracking methods, there are still some differences between the EM perturbations in the laboratory frame system and the ones in the standard GW frame system. In the standard GW frame system, the tensor polarizations ($h_\oplus$ and $h_\otimes$) of GWs satisfy the TT gauge condition. Therefore, studying the observable quantities of the EM counterparts (i.e., the PPFs) generated by HFGWs in the laboratory frame system is an important issue.

\subsection{Observable quantities of the EM counterparts in the laboratory frame system}	

In curved spacetime, only the local measurement made by an observer traveling in his/her world line has definite observable meaning. The corresponding observable quantities are just the projections of the physical quantities as a tensor on tetrads of the observer's world line. The tetrads consist of three mutually orthogonal space-like vectors and a time-like vector directed along the four-velocity of the observer. The time-like vector is perpendicular to the space-like vectors. We indicate the tetrads with $\tau^{\mu}_{(\alpha)}$, where the lower index in parentheses numbers the vectors and the upper index refers to components of the tetrads in the chosen frame. Consequently, the quantities $F_{(\alpha\beta)}$ measured by the observer are the tetrad components of the EM field tensor, which are given as
	\begin{align}\label{new18}
		F_{(\alpha\beta)}=F_{(\mu\nu)}\tau^{\mu}_{(\alpha)}\tau^{\nu}_{(\beta)}.
	\end{align}
For the 3DSR system, the observer should be at rest in the static magnetic field $\hat B^{(0)}$, so only the zeroth component of the four-velocity is non-vanishing. Therefore, the tetrad $\tau^{\nu}_{(0)}$ is given by
	\begin{align}\label{newb20}
		\tau^{\nu}_{(0)}=(\tau^{0}_{(0)},0,0,0).
	\end{align}
Using Eqs.~(\ref{1}) to (\ref{3}) and the orthonormality of the tetrads $g_{\mu\nu}\tau^{\mu}_{(\alpha)}\tau^{\nu}_{(\beta)}=\eta_{\alpha\beta}$, neglecting the higher order infinitesimals ($\sim h^2$), we find that the tetrads $\tau^{\mu}_{(\alpha)}$ in the HFGW field having additional polarization states [see Eqs.~(\ref{1}), (\ref{2}), (\ref{newb3}) and (\ref{4})] can be expressed as
	\begin{align}\label{newb21}
		&\tau_{(0)}^{\mu}=[1,0,0,0], \nonumber\\
		&\tau_{(1)}^\mu=\left[0,1-\frac{1}{2} h_{11}, 0,-h_{13}\right], \nonumber \\
		&	\tau_{(2)}^\mu=\left[0,-h_{12}, 1-\frac{1}{2} h_{22},-h_{23}\right], \nonumber \\
		&	\tau_{(3)}^\mu=\left[0,0,0,1-\frac{1}{2} h_{33}\right].
	\end{align}
Here, $h_{11}, h_{22}, h_{33}, h_{12}, h_{13}$, and $h_{23}$ are given by Eq.~(\ref{4}). Obviously, once $h_x=h_y=h_b=h_l=0$ and $\phi=0$ in Eqs.~(\ref{4}), (\ref{7}) to (\ref{10}), then Eq.~(\ref{newb21}) will be reduced to the expression of the tetrads $\tau_{(\alpha)}^\mu$ in the standard GW frame system (the TT frame system)~\cite{Li:2003tv}. Moreover, Eq.~(\ref{newb21}) indicates that the zeroth component of $\tau_{(\alpha)}^\mu$ coincides completely with the time axis in the standard GW frame system, while the deviations of $\tau_{(1)}^\mu, \tau_{(2)}^\mu,$ $\tau_{(3)}^\mu$ from the $x$-, $y$-, $z$- axes in the standard GW frame system are only on the order of $h$.
	
In this case, using Eqs.~(\ref{new18}) and (\ref{newb21}), considering $\left|\tilde{E}^{(0)}\right| \ll\left|\hat{E}^{(0)}\right|$ and $\left|\tilde{B}^{(0)}\right| \ll\left|\hat{B}^{(0)}\right|$, and neglecting the higher order infinitesimals ($\sim h^2$), one can obtain the observable quantities of the pure perturbative EM fields $\tilde F^{(1)}_{\mu\nu}$ in the tetrad system as follows:
	\begin{align}\label{newb22}
		&\tilde{E}_{(x)}^{(1)}=-c\tilde F_{(01)}^{(1)}=-c\tilde F_{\mu\nu}^{(1)}\tau^{\mu}_{(0)}\tau^{\nu}_{(1)}
		=\tilde{E}_x^{(1)}-\frac{1}{2} h_{11} \hat E_x^{(0)}-h_{13} \hat{E}_z^{(0)}, \nonumber\\
		&\tilde{E}_{(y)}^{(1)}=-c\tilde F_{(02)}^{(1)}=-c\tilde F_{\mu\nu}^{(1)}\tau^{\mu}_{(0)}\tau^{\nu}_{(2)}
		=\tilde{E}_y^{(1)}- h_{12} \hat E_x^{(0)}-\frac{1}{2} h_{22} \hat E_y^{(0)}-h_{23} \hat{E}_z^{(0)},\nonumber\\
		&\tilde{E}_{(z)}^{(1)}=-c\tilde F_{(03)}^{(1)}=-c\tilde F_{\mu\nu}^{(1)}\tau^{\mu}_{(0)}\tau^{\nu}_{(3)}
		=\tilde{E}_z^{(1)}-\frac12h_{33} \hat{E}_z^{(0)},
	\end{align}
	\begin{align}\label{newb23}
		&\tilde{B}_{(x)}^{(1)}=\tilde F_{(23)}^{(1)}= F_{\mu\nu}\tau^{\mu}_{(2)}\tau^{\nu}_{(3)}
		=\tilde{B}_x^{(1)}-\frac{1}{2} (h_{22}+h_{23}) \hat B_x^{(0)}+h_{12} \hat{B}_y^{(0)}, \nonumber\\
		&\tilde{B}_{(y)}^{(1)}=-\tilde F_{(13)}^{(1)}= -F_{\mu\nu}\tau^{\mu}_{(1)}\tau^{\nu}_{(3)}
		=\tilde{B}_y^{(1)}-\frac{1}{2} (h_{11}+h_{33}) \hat B_y^{(0)}, \nonumber\\
		&\tilde{B}_{(z)}^{(1)}=\tilde F_{(12)}^{(1)}= F_{\mu\nu}\tau^{\mu}_{(1)}\tau^{\nu}_{(2)}
		=\tilde{B}_z^{(1)}+h_{13}\hat B_x^{(0)}+h_{23}\hat B_y^{(0)}-\frac{1}{2} (h_{11}+h_{22}) \hat B_z^{(0)}.
	\end{align}
Eqs.~(\ref{newb22}) and (\ref{newb23}) show that there exists EM response of the longitudinal static EM fields $\hat E_z^{(0)}$ and $\hat B_z^{(0)}$ to HFGWs in the tetrad system, while the longitudinal static EM fields have no such an effect in the standard GW frame system (the TT frame system of the GR framework). It is an important difference between the general tetrad system and the standard GW frame system.
	
In our 3DSR system, since $\hat E_x^{(0)}=\hat E_y^{(0)}=\hat E_z^{(0)}=0$, Eq.~(\ref{newb22})  can be reduced to
	\begin{align}\label{newb24}
		\tilde E_{(x)}^{(1)}=\tilde E_{x}^{(1)},~~~~\tilde E_{(y)}^{(1)}=\tilde E_{y}^{(1)},~~~~\tilde E_{(z)}^{(1)}=\tilde E_{z}^{(1)}.
	\end{align}
	
\subsubsection{PPFs in the standard GW frame system ($\phi=0,~\hat{x}=x,~\hat{y}=y,~\hat{z}=z$)}

In this section, we study the PPFs in the standard GW frame systems.
	
(1) Distinguishing and probing the pure $\otimes$-type polarization and the combined state of the $\oplus$-type and $l$-type polarizations of HFGWs.
	
Putting $\hat{B}^{(0)}$ along the $y$-direction, then the 3DSR system becomes a coupling system between the transverse static magnetic field $\hat{B}_y^{(0)}$ and the GB (see Fig.~\ref{F3}). In this case, we have $\hat{B}_x^{(0)}=\hat{B}_z^{(0)}=0$ and $\hat{B}_y^{(0)}\not=0$. From Eqs.~(\ref{7}), (\ref{9}) and (\ref{newb22}), we can obtain
	\begin{align}\label{21}
		\tilde{E}_{(x)}^{(1)}&=\frac{i}{2} k_{g} c z\left(h_{\oplus}+\frac{1}{2} h_{l}\right) \hat{B}_{y}^{(0)},\\\label{22}
		\tilde{E}_{(y)}^{(1)}&=-\frac{i}{2} k_{g} c zh_{\otimes} \hat{B}_{y}^{(0)}.
	\end{align}
According to Eqs.~(\ref{21}), (\ref{22}) and (\ref{new14}), the PPFs propagating along the $\hat x$-direction and $\hat y$-direction are, respectively,
	\begin{align}\label{23}
n_{\hat x-\otimes}^{(1)}
		=& \frac{c}{2 \hbar \omega_{e}}\operatorname{Re}\langle \mathop{T}^{(1)}\!^{01}\rangle_{\omega_{e}=\omega_{g}}       \notag\\
		=&\frac{1}{2 \mu_{0} \hbar \omega_{e}} \operatorname{Re}\left\langle\tilde{E}_{(y)}^{(1) *} \tilde{B}_{(z)}^{(0)}\right\rangle_{\omega_{e}=\omega_{g}}                 \notag\\
		=&\frac{A_{\otimes} \hat{B}_{y}^{(0)} \psi_{0}  \hat z }{4 \mu_{0} \hbar \sqrt{1+(z / f)^{2}}} \exp \left(-\frac{r^{2}}{W^{2}}\right)                           \notag\\
		& \sin \left[\frac{k_g r^{2}}{2R} -\tan ^{-1} \left(\frac{z}{f}\right)+\delta\right]            \notag\\
		=&\frac{A_{\otimes} \hat{B}_{y}^{(0)} \psi_{0}  \hat z }{4 \mu_{0} \hbar}
		\mathscr{A}\sin \mathscr{B},\\\label{24}
n_{\hat y-\oplus,l}^{(1)}
		=& \frac{c}{2 \hbar \omega_{e}}\operatorname{Re}\langle \mathop{T}^{(1)}\!^{02}\rangle_{\omega_{e}=\omega_{g}}       \notag\\
		=&\frac{-1}{2 \mu_{0} \hbar \omega_{e}} \operatorname{Re}\left\langle\tilde{E}_{(x)}^{(1) *} \tilde{B}_{(z)}^{(0)}\right\rangle_{\omega_{e}=\omega_{g}}                       \notag\\
		=&\frac{\left(A_{\oplus}+\sqrt{2} / 2 A_l\right)  \hat{B}_{y}^{(0)} \psi_{0}   \hat z }{4 \mu_{0} \hbar \sqrt{1+(z / f)^{2}}} \exp \left(-\frac{r^{2}}{W^{2}}\right)     \notag\\
		&\sin \left[\frac{k_g r^{2}}{2R} -\tan ^{-1} \left(\frac{z}{f}\right)+\delta\right]            \notag\\
		=& \frac{\left(A_{\oplus
			}+\frac{\sqrt{2}}{2}A_l\right) \hat{B}_{y}^{(0)} \psi_{0}  \hat z }{4 \mu_{0} \hbar}
		\mathscr{A}\sin \mathscr{B},
	\end{align}
where
	\begin{align}\label{25}
		\mathscr{A}=&\frac{\exp \left(-\frac{r^{2}}{W^{2}}\right)}{\sqrt{1+(z / f)^{2}}}, \\\label{26}
		\mathscr{B}=&\frac{k_g r^{2}}{2R} -\tan ^{-1} \left(\frac{z}{f}\right)+\delta.
	\end{align}
Note that $n_{\hat x-\otimes}^{(1)}$ is the PPF propagating along the $\hat x$-direction generated by the pure $\otimes$-type polarization of HFGWs, while $n_{\hat y-\oplus,l}^{(1)}$ is the PPF propagating along the $\hat y$-direction produced by the combined state of the $\oplus$-type and $l$-type polarizations of HFGWs (see Fig.~\ref{F3}). The parameters $\mathop{T} \limits^{(1)}\!^{01}$ and  $\mathop{T} \limits^{(1)}\!^{02}$ are $01$ and $02$ components of the energy-momentum tensors of the first-order perturbative EM fields, respectively. The variable $\hat z$ is the propagating distance of HFGWs in the static magnetic field of the 3DSR system, which shows a spatial accumulation effect~\cite{Li:2017jcz,Boccaletti1970} for HFGWs in the EM response.

(2) Distinguishing and probing the combined state of the $\oplus$-type, $b$-type and $l$-type  polarizations and the pure $\otimes$-type polarization of HFGWs.
	
Putting $\hat{B}^{(0)}$ along the $\hat x$-direction, then the 3DSR system becomes a coupling system between the transverse static magnetic field $\hat{B}_x^{(0)}$ and the GB (see Fig.~\ref{F4}). In this case, we have $\hat{B}_y^{(0)}=\hat{B}_z^{(0)}=0$ and $\hat{B}_x^{(0)}\not=0$. From Eqs.~(\ref{7}), (\ref{9}) and (\ref{newb24}), we can get
\begin{align}\label{27}
		\tilde{E}_{(x)}^{(1)}=&-\frac{i}{2} k_{g} c \hat z h_{\otimes} \hat{B}_{x}^{(0)},\\\label{28}
		\tilde{E}_{(y)}^{(1)}=&\frac{i}{2} k_{g} c \hat z\left(h_b-h_{\oplus}+ h_{l}\right) \hat{B}_{x}^{(0)}.
	\end{align}
In the same way, the PPFs propagating along the $\hat{x}$-direction and $\hat{y}$-direction are given by, respectively,
	\begin{align}\label{29}
n_{\hat x-\oplus,b,l}^{(1)}
		=& \frac{c}{2 \hbar \omega_{e}}\operatorname{Re}\langle \mathop{T}^{(1)}\!^{01}\rangle _{\omega_{e}=\omega_{g}}       \notag\\
		=&\frac{1}{2 \mu_{0} \hbar \omega_{e}} \operatorname{Re}\left\langle\tilde{E}_{(y)}^{(1) *} \tilde{B}_{(z)}^{(0)}\right\rangle_{\omega_{e}=\omega_{g}}                 \notag\\
		=&\frac{\left( A_b -A_{\oplus}+ \sqrt{2} A_l\right) \hat{B}_{(x)}^{(0)} \psi_{0}  \hat z }{4 \mu_{0} \hbar }\mathscr{A}\sin \mathscr{B} ,\\\label{30}
n_{\hat y-\otimes}^{(1)}
		=& \frac{c}{2 \hbar \omega_{e}}\operatorname{Re}\langle \mathop{T}^{(1)}\!^{02}\rangle _{\omega_{e}=\omega_{g}}       \notag\\
		=&\frac{-1}{2 \mu_{0} \hbar \omega_{e}} \operatorname{Re}\left\langle\tilde{E}_{(x)}^{(1) *} \tilde{B}_{(z)}^{(0)}\right\rangle_{\omega_{e}=\omega_{g}}                 \notag\\
		=&\frac{ A_{\otimes} \hat{B}_{(x)}^{(0)} \psi_{0}  \hat z }{4 \mu_{0} \hbar }\mathscr{A}\sin \mathscr{B},
	\end{align}	
where $n_{\hat x-\oplus,b,l}^{(1)}$ is the PPF propagating along the $\hat x$-direction generated by the combined state of the $\oplus$-type, $b$-type and $l$-type polarizations of HFGWs, and $n_{\hat y-\otimes}^{(1)}$ is the PPF propagating along the $\hat y$-direction produced by the pure $\otimes$-type polarization of HFGWs (see Fig.~\ref{F4}).
	
(3) Distinguishing and probing the pure $x$-type polarization and the pure $y$-type polarization of HFGWs.
	
Putting $\hat{B}_{}^{(0)}$ along the $\hat{z}$-direction, then the 3DSR system becomes a coupling system between the longitude static magnetic field $\hat{B}_{z}^{(0)}$ and the GB (see Fig.~\ref{F5}). In this case, we have $\hat{B}_{y}^{(0)}=\hat{B}_{x}^{(0)}=0$ and $\hat{B}_{z}^{(0)}\not=0$. From  Eqs.~(\ref{7}), (\ref{9}) and (\ref{newb24}), we can get
	\begin{align}\label{31}
		\tilde{E}_{(x)}^{(1)}=&-\frac{i}{2} k_{g} c \hat z h_y \hat{B}_{z}^{(0)},\\\label{32}
		\tilde{E}_{(y)}^{(1)}=&\frac{i}{2} k_{g} c \hat z h_x \hat{B}_{z}^{(0)}.
	\end{align}			
Then, the PPFs propagating along the $\hat x$-direction and $\hat y$-direction are given by, respectively,
	\begin{align}\label{33}
n_{\hat x-x}^{(1)}
		=& \frac{c}{2 \hbar \omega_{e}}\operatorname{Re}\langle \mathop{T}^{(1)}\!^{01}\rangle_{\omega_{e}=\omega_{g}}       \notag\\
		=&\frac{1}{2 \mu_{0} \hbar \omega_{e}} \operatorname{Re}\left\langle\tilde{E}_{(y)}^{(1) *} \tilde{B}_{(z)}^{(0)}\right\rangle_{\omega_{e}=\omega_{g}}                 \notag\\
		=&\frac{ A_x \hat{B}_{z}^{(0)} \psi_{0}   \hat z }{4 \mu_{0} \hbar }\mathscr{A}\sin \mathscr{B} ,\\
		\notag\label{34}
n_{\hat y-y}^{(1)}
		=& \frac{c}{2 \hbar \omega_{e}}\operatorname{Re}\langle \mathop{T}^{(1)}\!^{02}\rangle_{\omega_{e}=\omega_{g}}       \notag\\
		=&\frac{-1}{2 \mu_{0} \hbar \omega_{e}} \operatorname{Re}\left\langle\tilde{E}_{(x)}^{(1) *} \tilde{B}_{(z)}^{(0)}\right\rangle_{\omega_{e}=\omega_{g}}                 \notag\\
		=&\frac{ A_y \hat{B}_{z}^{(0)} \psi_{0}  \hat z }{4 \mu_{0} \hbar}\mathscr{A}\sin \mathscr{B},
	\end{align}
where $n_{\hat x-x}^{(1)}$ is the PPF propagating along the $\hat x$-direction generated by the pure $x$-type polarization of HFGWs, and $n_{\hat y-y}^{(1)}$ is the PPF propagating along the $\hat y$-direction produced by the pure $y$-type polarization of HFGWs (see Fig.~\ref{F5}). It is worth mentioning that $n_{\hat x-x}^{(1)}$ and $n_{\hat y-y}^{(1)}$ are only produced by the pure additional polarization states (the $x$-type and $y$-type polarizations, i.e., the vector-mode gravitons) of HFGWs. It is a new characteristic which cannot be found in the GR framework, and so it could provide an effective way to directly test the gravitational theorise beyond the GR.

It is interesting to compare the PPF $n^{(1)}_{\hat x-\otimes}$ [see Eq.~(\ref{23})] and the background noise photon flow (BPF) resulted from the GB. In fact, the BPF resulted from the GB is the dominant noise photon flow in the 3DSR system because it is much larger than other noise photon flows, such as shot noise, thermal noise (operating temperature lower than $1\,$K), Johnson noise, quantum noise, etc.~\cite{Woods2011}. Therefore, once the PPF $n^{(1)}_{\hat x-\otimes}$ can be distinguished and identified from the BPF resulted from the GB, the PPF $n^{(1)}_{\hat x-\otimes}$ (and the other PPFs) would also be distinguished and identified from the other noise photons.

Reviewing Eqs.~(\ref{new14}) and (\ref{new16}), we have
	\begin{align}\label{new35}
		\tilde E^{(0)}_{(y)}=\psi _{ey}=\frac{2cx}{k_e}\left(\frac{i}{W^2}+\frac{k_e}{2R}\right)\psi.
	\end{align}
Thus, the BPF propagating along the $\hat x$-direction can be given by
	\begin{align}\label{new36}
		n^{(0)}_{\hat x}&=\frac{1}{2 \mu_{0} \hbar \omega_{e}} \operatorname{Re}\left\langle\tilde{E}_{(y)}^{(0) *} \tilde{B}_{(z)}^{(0)}\right\rangle_{\omega_{e}=\omega_{g}} \nonumber\\
		&=\frac{  \psi^2_{0}  x }{2\mu_{0} \hbar k_e R\left[{1+(z / f)^{2}}\right] }\text{exp}\left(-\frac{2r^2}{W^2}\right).
	\end{align}
Comparing Eqs.~(\ref{23}) and~(\ref{new36}), one can find the following important properties:
	
(a) $n^{(1)}_{\hat x-\otimes}$ is an even function of the coordinates $x$, which means that $n^{(1)}_{\hat x-\otimes}$ has the same propagating direction in the regions of $x>0$ and $x<0$ (see Fig.~\ref{F7}). However, $n^{(0)}_{\hat x}$ is an odd function of the coordinates $x$, so the propagating directions of $n^{(0)}_{\hat x}$ are opposite in the regions of $x>0$ and $x<0$ (see Fig.~\ref{F6}).
	
(b) Eqs.~(\ref{23}) and (\ref{new36}) indicate
	\begin{align}
		n_{\hat x-\otimes}^{(1)}\propto\frac{\exp \left(-\frac{r^{2}}{W^{2}}\right)}{\sqrt{1+(z/f)^2}}=\frac{\exp \left(-\frac{x^{2}+y^{2}}{W^{2}}\right)}{\sqrt{1+(z/f)^2}},
		\\
		n_{\hat x}^{(0)}\propto\frac{\exp \left(-\frac{2 r^{2}}{W^{2}}\right)}{{1+(z/f)^2}}=\frac{\exp \left(-\frac{2\left(x^{2}+y^{2}\right)}{W^{2}}\right)}{{1+(z/f)^2}}.
	\end{align}
Therefore, the decay rate of $n_{\hat x-\otimes}^{(1)}$ is observably slower than that of $n_{\hat x}^{(0)}$, although the peak value of $n_{\hat x}^{(0)}$ is much larger than that of $n_{\hat x-\otimes}^{(1)}$. However, it is always possible to find suitable positions where $n_{ \hat x-\otimes}^{(1)}$ and $n_{\hat x}^{(0)}$ are comparable and distinguishable due to their different physical behaviors such as peak value positions, propagating directions, decay rate, wave impedance, etc. Especially, the signal photon flows generated by some HFGWs could have large spectral densities (e.g., the HFGWs expected by the brane oscillation~\cite{Clarkson:2006pq,Seahra:2004fg}, the evaporation of primordial black holes~\cite{Aggarwal:2020olq}, the interaction between astrophysical plasma and intense EM radiation~\cite{Servin:2003cf}, etc., see Fig.~\ref{F8} and Table~\ref{tab1}), which makes it quite possible  to distinguish the PPF $n^{(1)}_{\hat x-\otimes}$ from the BPF $n_{\hat x}^{(0)}$. By the way, for the electromagnetic waves (photon flows) in the GHz band, the wave impedances of copper, silver and gold are $0.060\,\Omega$, $0.063\,\Omega$ and $0.046\,\Omega$, respectively, and the usual superconductor is about $10^{-3}\,\Omega$ to them~\cite{Haslett}. The wave impedance of the 3DSR system to the PPFs is only $10^{-4}\,\Omega$ or less~\cite{Li:2015tqa}. Therefore, the 3DSR system looks like a ``good equivalent superconductor'' to the PPFs. Moreover, the wave impedance to the BPF resulted from the GB and other noise photons in the 3DSR system is about 300 $\Omega$ or larger~\cite{Li:2015tqa}. Therefore, wave impedance matching will be an effective way to distinguish the PPFs and the BPFs in the 3DSR system.

\subsubsection{PPFs in the non-standard GW frame systems ($\phi=\frac{\pi}4$ and $\phi=\frac{\pi}2$)}
	
Next, we study the PPFs in the non-standard GW frame systems. We consider two special cases: $\phi=\frac{\pi}4$ and $\phi=\frac{\pi}2$.
	
(1)  $\phi=\frac{\pi}{4}$: distinguishing and probing the pure $\oplus$-type polarization and  combined state of the $\otimes$-type and $l$-type polarizations of HFGWs.

When $\phi=\frac{\pi}{4}$, from Eq.~(\ref{4}), we have
	\begin{align}\label{35}
		{h}_{11}=& h_{\otimes}+h_b,                     \notag   \\
		{h}_{12}=& {h}_{21}= -h_{\oplus},              \notag    \\
		{h}_{22}=& -h_{\otimes}+h_b,                       \notag       \\
		{h}_{13}=& {h}_{31} = \frac{\sqrt{2}}{2}\left(h_{x}+h_y\right),\notag \\
		{h}_{23}=& {h}_{32} = \frac{\sqrt{2}}{2}\left(-h_{x}+h_y\right),\notag \\
		\hat{h}_{33}=&h_l.
	\end{align}	
Comparing {Eqs.~(\ref{newb3})} and (\ref{35}), it can be seen that the positions of the $\oplus$-type and $\otimes$-type polarizations are exchanged to each other.

Putting $\hat B^{(0)}$ along the $\hat y$ direction, then the EM response to HFGWs in the 3DSR system could be a coupling system between the transverse static magnetic field  $\hat{B}_y^{(0)}$ and the GB. In this case, we have $\hat{B}_x^{(0)}=\hat{B}_z^{(0)}=0$ and $\hat{B}_y^{(0)} \not= 0$. According to Eqs.~(\ref{7}), (\ref{9}) and (\ref{newb22}), one can obtain
	\begin{align}\label{xx1}
		\tilde{E}_{(x)}^{(1)}=&\frac{i}{2} k_{g} c \hat z \left(h_\otimes + \frac12 h_l \right)  \hat{B}_{y}^{(0)},\\\label{xx2}
		\tilde{E}_{(y)}^{(1)}=&-\frac{i}{2} k_{g} c \hat z h_\oplus\hat{B}_{y}^{(0)}.
	\end{align}
Combining with Eqs.~(\ref{7}), (\ref{9}), (\ref{xx1}) and (\ref{xx2}), the PPFs propagating along the $\hat x$-direction and $\hat y$-direction are given as, respectively,
	\begin{align}\label{36}
n_{\hat x-\oplus}^{(1)}
		=& \frac{c}{2 \hbar \omega_{e}}\operatorname{Re}\langle \mathop{T}^{(1)}\!^{01}\rangle_{\omega_{e}=\omega_{g}}      \notag\\
		=&\frac{1}{2 \mu_{0} \hbar \omega_{e}} \operatorname{Re}\left\langle\tilde{E}_{(y)}^{(1) *} \tilde{B}_{(z)}^{(0)}\right\rangle_{\omega_{e}=\omega_{g}}      \notag\\
		=&\frac{-A_{\oplus}  \hat{B}_{y}^{(0)} \psi_{0}  \hat z }{4 \mu_{0} \hbar }\mathscr{A}\sin \mathscr{B} , \\
		\notag\label{37}
n_{\hat y-\otimes,l}^{(1)}
		=& \frac{c}{2 \hbar \omega_{e}}\operatorname{Re}\langle \mathop{T}^{(1)}\!^{02}\rangle_{\omega_{e}=\omega_{g}}  \notag\\
		=&\frac{-1}{2 \mu_{0} \hbar \omega_{e}} \operatorname{Re}\left\langle\tilde{E}_{(x)}^{(1) *} \tilde{B}_{(z)}^{(0)}\right\rangle_{\omega_{e}=\omega_{g}}                 \notag\\
		=&\frac{ -\left(A_{\otimes} + \sqrt{2} / 2A_l\right) \hat{B}_{y}^{(0)} \psi_{0}  \hat z }{4 \mu_{0} \hbar }\mathscr{A}\sin \mathscr{B},
	\end{align}
where $n_{\hat x-\oplus}^{(1)}$ is the PPF propagating along the $\hat{x}$-direction generated by the pure $\oplus$-type polarization of HFGWs, and $n_{\hat y-\otimes,l}^{(1)}$ is the PPF propagating along the $\hat{y}$-direction produced by the combined state of the $\otimes$-type and $l$-type polarizations of HFGWs.
	
The EM response to HFGWs in the 3DSR system could also be a coupling system between the longitude static magnetic field $\hat B_z^{(0)}$ and the GB. Then, we have $\hat B_x^{(0)}=\hat B_y^{(0)}=0$ and $\hat B_z^{(0)}\neq0$. From Eqs.~(\ref{7}), (\ref{9}) and (\ref{newb24}), we can obtain
	\begin{align}\label{38}
		\tilde E_{(x)}^{(1)}=i \frac{\sqrt{2}}{4}k_g c\hat z[h_x-h_y]\hat B_z^{(0)},\\\label{39}
		\tilde E_{(y)}^{(1)}=i \frac{\sqrt{2}}{4}k_g c\hat z[h_x+h_y]\hat B_z^{(0)}.
	\end{align}
Clearly, if $h_x=h_y$, they can be reduced to
	\begin{align}\label{40}
		\tilde E_{(x)}^{(1)}=\,\,&0,\\\label{41}
		\tilde E_{(y)}^{(1)}=\,\,&i \frac{\sqrt{2}}{2}k_g c\hat zh_x\hat B_z^{(0)}
		=i \frac{\sqrt{2}}{2}k_g c\hat zh_y\hat B_z^{(0)}.
	\end{align}
In this case, we have
	\begin{align}\label{42}
		n_{\hat x-x}^{(1)}=n_{\hat x-y}^{(1)}=&\frac{c}{2\hbar \omega_{e}} \operatorname{Re}\langle \mathop{T}^{(1)}\!^{01}\rangle_{\omega_{e}=\omega_{g}}\notag\\
		=&\frac{1}{2 \mu_{0} \hbar \omega_{e}} \operatorname{Re}\left\langle\tilde{E}_{(y)}^{(1) *} \tilde{B}_{(z)}^{(0)}\right\rangle_{\omega_{e}=\omega_{g}}                 \notag\\
		=&\frac{\sqrt{2} A_x \hat{B}_{(z)}^{(0)} \psi_{0}  \hat z }{4 \mu_{0} \hbar }\mathscr{A}\sin \mathscr{B}
		\notag\\
		=&\frac{\sqrt{2} A_y \hat{B}_{(z)}^{(0)} \psi_{0}  \hat z }{4 \mu_{0} \hbar }\mathscr{A}\sin \mathscr{B},\\\label{43}
		n_{\hat y-x}^{(1)}=n_{\hat y-y}^{(1)}=&0.
	\end{align}
Eqs.~(\ref{42}) and (\ref{43}) mean that there is only the PPF propagating along the $\hat x$-direction and no the PPF propagating along the $\hat y$-direction. The PPF propagating along the $\hat x$-direction is the sum of contribution from the $x$-type and $y$-type polarizations of HFGWs.
	
(2) $\phi=\frac{\pi}{2} $: distinguishing and probing the pure $\otimes$-type polarization and combined state of the $\oplus$-type and $l$-type polarizations of HFGWs.
	
When $\phi=\frac{\pi}{2} $, from Eq.~(\ref{4}), we have
	\begin{align}\label{44}
		{h}_{11}= & -h_{\oplus}+h_b,                     \notag      \\
		{h}_{12}= & {h}_{21}= -h_{\otimes},              \notag  \\
		{h}_{22}= & h_{\oplus}+h_b,                       \notag    \\
		{h}_{13}= & {h}_{31} = h_y                        \notag \\
		{h}_{23}= & {h}_{32} = -h_x,\notag                        \\
		{h}_{33}= &  h_l.
	\end{align}
	
In this case, the EM response to HFGW in the 3DSR system could be a coupling system between the transverse static magnetic field  $\hat{B}_y^{(0)}$ and the GB. Then, we have $\hat{B}_x^{(0)}=\hat{B}_z^{(0)}=0$ and $\hat{B}_y^{(0)} \not= 0$. In the same way, from Eqs.~(\ref{7}), (\ref{9}) and (\ref{newb22}), we can obtain
	\begin{align}\label{45}
		\tilde{E}_{(x)}^{(1)}=&-\frac{i}{2} k_{g} c \hat z \left(-h_\oplus + \frac12 h_l \right)  \hat{B}_{(y)}^{(0)},\notag\\
		\tilde{E}_{(y)}^{(1)}=&-\frac{i}{2} k_{g} c \hat z h_\otimes\hat{B}_{(y)}^{(0)}.
	\end{align}
Then, the PPFs propagating along the $\hat x$-direction and $\hat y$-direction are given by, respectively,
	\begin{align}\label{46}
n_{\hat x-\otimes}^{(1)}
		=& \frac{c}{2 \hbar \omega_{e}}\operatorname{Re}\langle \mathop{T}^{(1)}\!\,^{01}\rangle_{\omega_{e}=\omega_{g}}
		\notag\\
		=&\frac{1}{2 \mu_{0} \hbar \omega_{e}} \operatorname{Re}\left\langle\tilde{E}_{(y)}^{(1) *} \tilde{B}_{(z)}^{(0)}\right\rangle_{\omega_{e}=\omega_{g}}      \notag\\
		=&\frac{- A_{\otimes}  \hat{B}_{(y)}^{(0)} \psi_{0} \hat  z }{4 \mu_{0} \hbar }\mathscr{A}\sin \mathscr{B} , \\\label{47}
n_{\hat y-\oplus,l}^{(1)}
		=& \frac{c}{2 \hbar \omega_{e}}\operatorname{Re}\langle \mathop{T}^{(1)}\!\,^{02}\rangle_{\omega_{e}=\omega_{g}}  \notag\\
		=&\frac{-1}{2 \mu_{0} \hbar \omega_{e}} \operatorname{Re}\left\langle\tilde{E}_{(x)}^{(1) *} \tilde{B}_{(z)}^{(0)}\right\rangle_{\omega_{e}=\omega_{g}}                 \notag\\
		=&\frac{ \left(-A_{\oplus} + \sqrt{2} / 2A_l\right) \hat{B}_{(y)}^{(0)} \psi_{0}  \hat z }{4 \mu_{0} \hbar }\mathscr{A}\sin \mathscr{B},
	\end{align}
where $n_{\hat x-\otimes}^{(1)}$ is the PPF propagating along the $\hat x$-direction generated by the pure $\otimes$-type polarization of HFGWs, and $n_{\hat y-\oplus,l}^{(1)}$ is the PPF propagating along the $\hat y$-direction produced by the combined state of the $\oplus$-type and $l$-type polarizations of HFGWs.
	
Note that although both Eqs.~(\ref{24}) and (\ref{47}) represent the transverse PPFs propagating along the $\hat{y}$-direction generated by the combined state of the $\oplus$-type and $l$-type polarizations of HFGWs, they correspond to different combination forms: the former is a ``constructive'' combination (i.e., the $\oplus$-type and $l$-type polarizations have the same symbols), and the latter is a ``destructive combination'' (i.e., the $\oplus$-type and $l$-type polarizations have opposite symbols). Therefore, according to Eqs.~(\ref{24}) and (\ref{47}), it can be found that
	\begin{align}\label{48}
		n_{\hat y-l}^{(1)}=&\left(n_{y-\otimes,l}^{(1)}\right)_{\phi=\frac{\pi}{2}}+\left(n_{y-\otimes,l}^{(1)}\right)_{\phi=0}\notag\\
		=&\frac{ \sqrt{2}A_l \hat{B}_{(y)}^{(0)} \psi_{0}  \hat z }{4 \mu_{0} \hbar }\mathscr{A}\sin \mathscr{B}.
	\end{align}	
So, $n_{\hat y-l}^{(1)}$ is just the value of the PPF propagating along the $\hat{y}$-direction generated by the pure $l$-type polarization of HFGWs.
	
The EM response to HFGWs in the 3DSR system could also be a coupling system between the longitude static magnetic field $\hat B_z^{(0)}$ and the GB. Then, we have $\hat B_x^{(0)}=\hat B_y^{(0)}=0$ and  $\hat B_z^{(0)}\neq 0$. From Eqs.~(\ref{7}), (\ref{9}) and (\ref{newb22}), we can obtain
	\begin{align}\label{49}
		\tilde E_{(x)}^{(1)}=\frac{{i}}{2}k_g c \hat z h_x\hat B_{(z)}^{(0)},\\\label{50}
		\tilde E_{(y)}^{(1)}=\frac{{i}}{2}k_g c \hat z h_y\hat B_{(z)}^{(0)}.
	\end{align}
Then, the PPFs propagating along the $\hat x$-direction and $\hat y$-direction are given by, respectively,
	\begin{align}\label{51}
	n_{\hat x-y}^{(1)}=&\frac{c}{2\hbar \omega_{e}} \operatorname{Re}\langle \mathop{T}^{(1)}\!^{01}\rangle_{\omega_{e}=\omega_{g}}\notag\\
		=&\frac{1}{2 \mu_{0} \hbar \omega_{e}} \operatorname{Re}\left\langle\tilde{E}_{(y)}^{(1) *} \tilde{B}_{(z)}^{(0)}\right\rangle_{\omega_{e}=\omega_{g}}                 \notag\\
		=&\frac{ A_y \hat{B}_{(z)}^{(0)} \psi_{0}  \hat z }{4 \mu_{0} \hbar }\mathscr{A}\sin \mathscr{B},\\\label{52}
	n_{\hat y-x}^{(1)}=&\frac{c}{2\hbar \omega_{e}} \operatorname{Re}\langle \mathop{T}^{(1)}\!^{02}\rangle_{\omega_{e}=\omega_{g}}\notag\\
		=&\frac{-1}{2 \mu_{0} \hbar \omega_{e}} \operatorname{Re}\left\langle\tilde{E}_{(x)}^{(1) *} \tilde{B}_{(z)}^{(0)}\right\rangle_{\omega_{e}=\omega_{g}}                 \notag\\
		=&\frac{ -A_x \hat{B}_{(z)}^{(0)} \psi_{0}  \hat z }{4 \mu_{0} \hbar }\mathscr{A}\sin \mathscr{B},
	\end{align}
where $n_{\hat x-y}^{(1)}$ is the PPF propagating along the  $\hat x$-direction generated by the pure $y$-type polarization of HFGWs, and $n_{\hat y-x}^{(1)}$ is the PPF propagating along the  $\hat y$-direction produced by the pure $x$-type polarization of HFGWs. It is interesting to compare Eqs.~(\ref{51}), (\ref{52}) with Eqs.~(\ref{33}), (\ref{34}). Obviously, the roles of the $x$-type and $y$-type polarizations for $\phi=0$ and $\phi=\frac{\pi}{2}$ exchange with each other. The propagating directions of the PPFs in the $\hat x$-direction for $\phi=0$ and $\phi=\frac{\pi}{2}$ are identical, while the propagating directions in the $\hat y$-direction are opposite.
	
For the stochastic and incoherent HFGWs (e.g., the primordial HFGWs), their distributions are almost uniform and isotropic. Thus, the intensities of the PPFs generated by these HFGWs are almost independent on the rotation of the 3DSR system. However, due to the randomness of the features of the HFGWs mentioned above (including their energy flows and phases, etc.), probing the stochastic and incoherent HFGWs will be more difficult than probing the coherent HFGWs with the same amplitudes. The intensities of the PPFs generated by the former will be one to two orders of magnitude smaller than those generated by the latter. But it is not impossible to probe the stochastic and incoherent HFGWs~\cite{Li:2009zzq,Tong:2009vk,XiuLin}.
	
Now, we have calculated the PPFs of various situations, including the PPFs generated by the pure $\otimes$-type polarization of HFGWs [see Eqs.~(\ref{23}), (\ref{30}) and (\ref{46})], the PPF generated by the pure $\oplus$-type polarization [see Eq.~(\ref{36})], the PPF generated by the pure $x$-type polarization [see Eq.~(\ref{33})], the PPF generated by the pure $y$-type polarization [see Eq.~(\ref{34})], and the PPF generated by the pure $l$-type polarization [see Eq.~(\ref{48})], respectively. Based on the results above, one can find that the pure $b$-type polarization of HFGWs is completely determined by Eq.~(\ref{29}). In principle, all the polarizations (i.e., the $\oplus$-type, $\otimes$-type, $x$-type, $y$-type, $l$-type, and $b$-type polarizations) of HFGWs can be distinguished and probed by the corresponding PPFs. Therefore, the six polarization states of HFGWs have separability and detectability in the 3DSR system.

\section{Numerical estimation: the conditions for displaying the PPFs in the BPF fluctuation}
	
\begin{figure}[H]
		\centering
		\renewcommand{\figurename}{Fig.}
		\includegraphics[width=15cm]{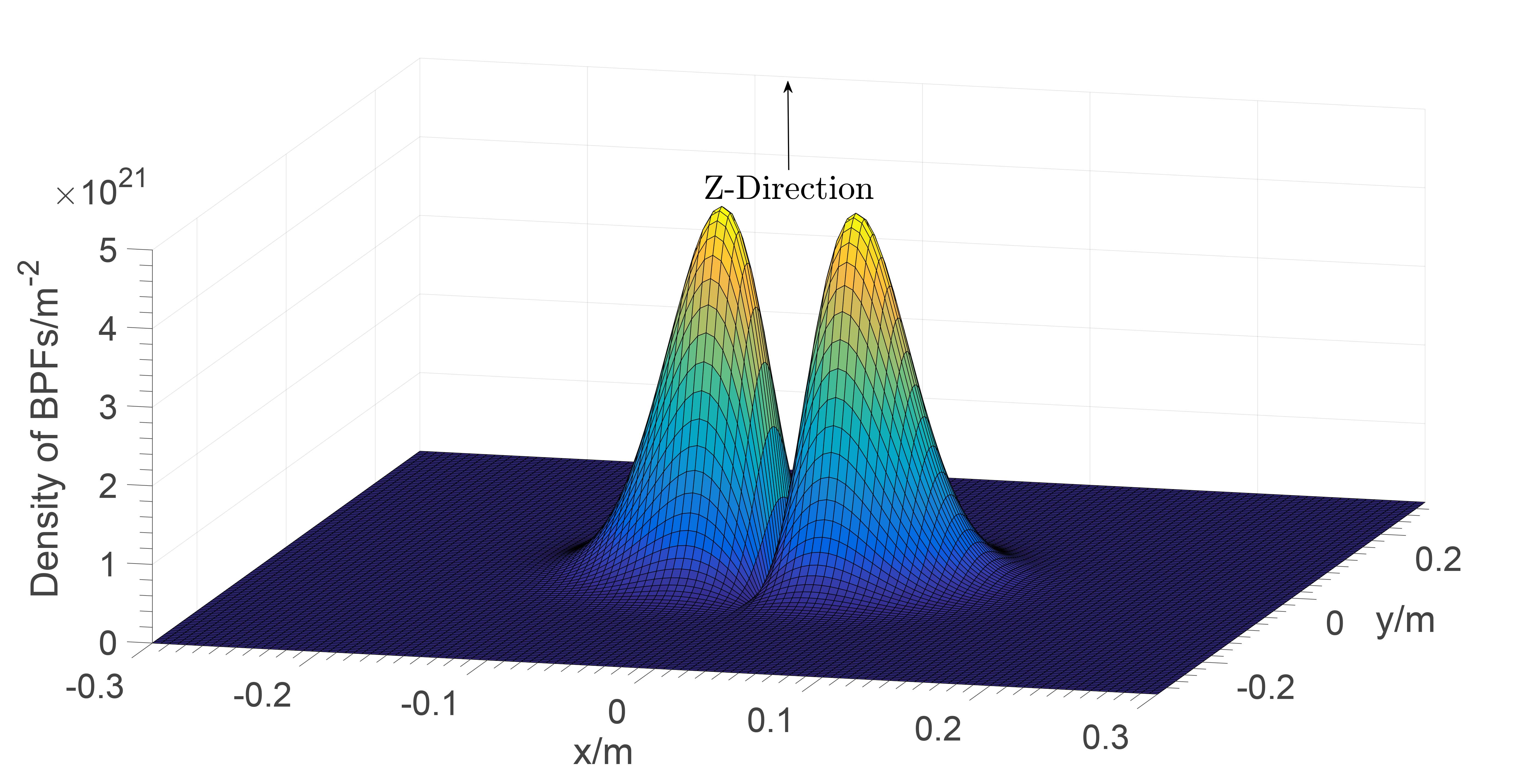}
		\includegraphics[width=15cm]{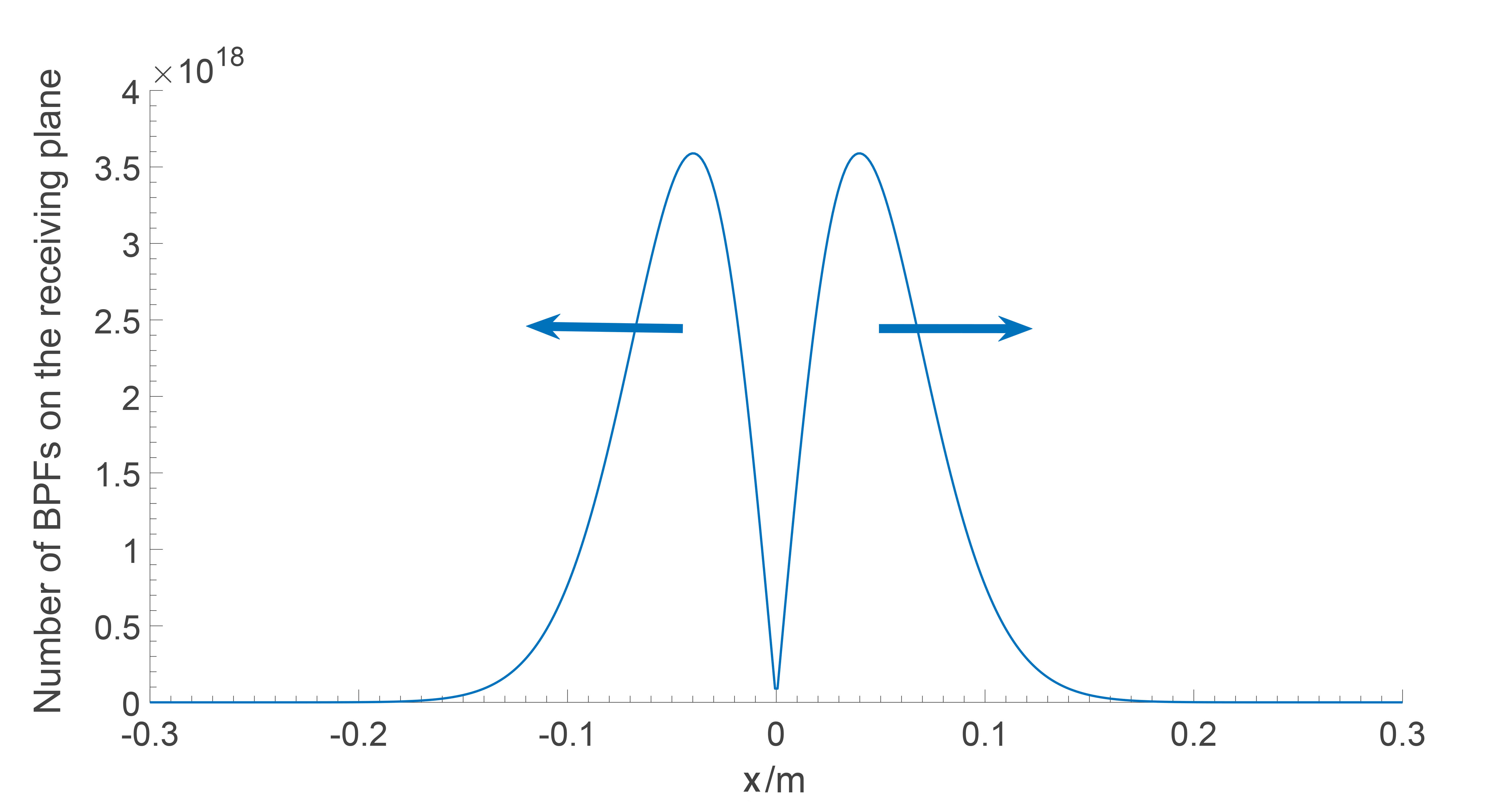}
		\caption{The transverse background noise photon flow $n_{\hat x}^{(0)}$ caused by the GB with $P=10$ W and $\nu=10$ GHz [see Eq.~(\ref{new36})] is an odd function of the coordinate $x$, so the propagating directions of $n_{\hat x}^{(0)}$ are opposite in the regions of $ x>0$ and $ x<0$.} \label{F6}		
\end{figure}

\begin{figure}[H]
		\centering
		\renewcommand{\figurename}{Fig.}	
		\includegraphics[width=15cm]{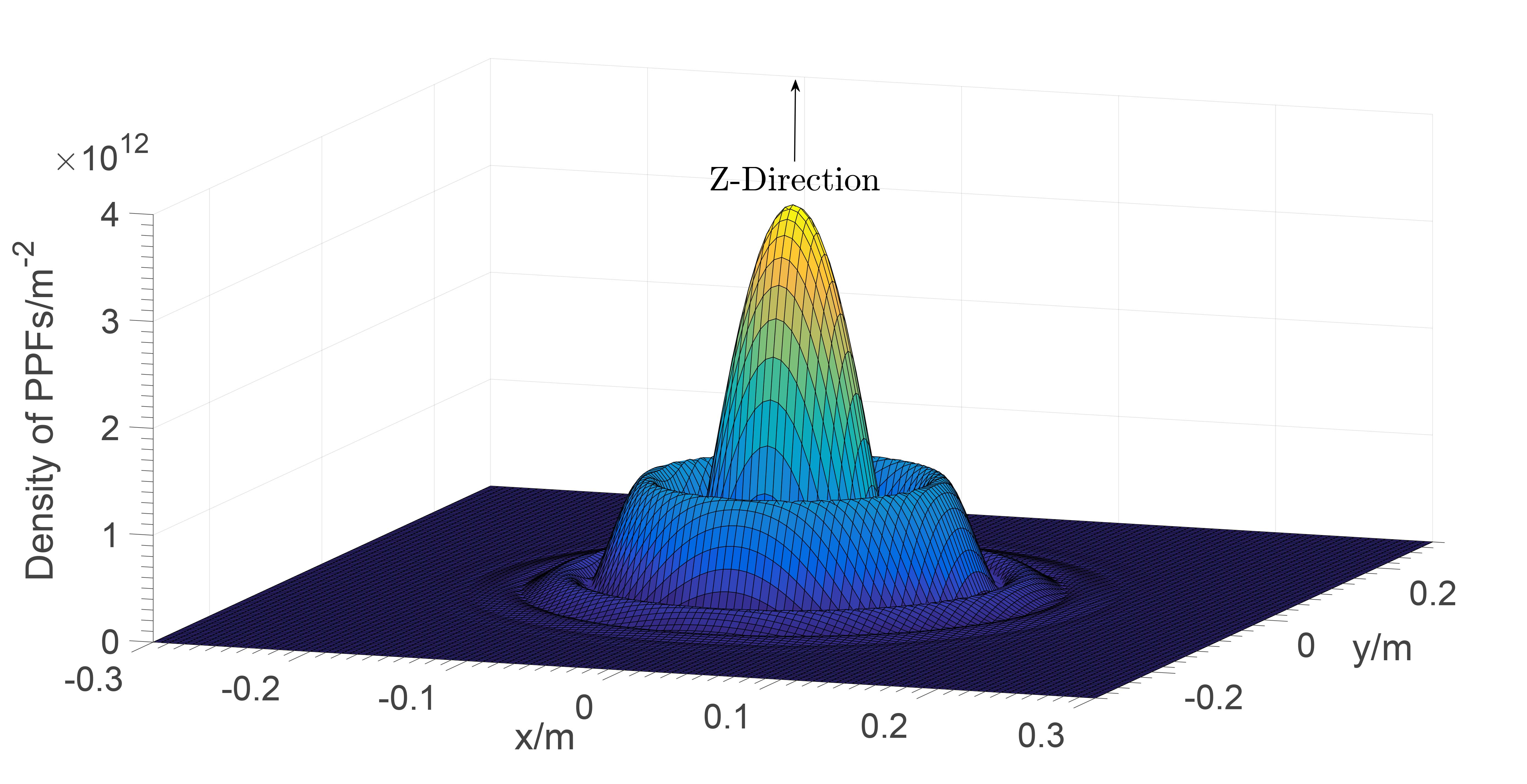}
		\includegraphics[width=15cm]{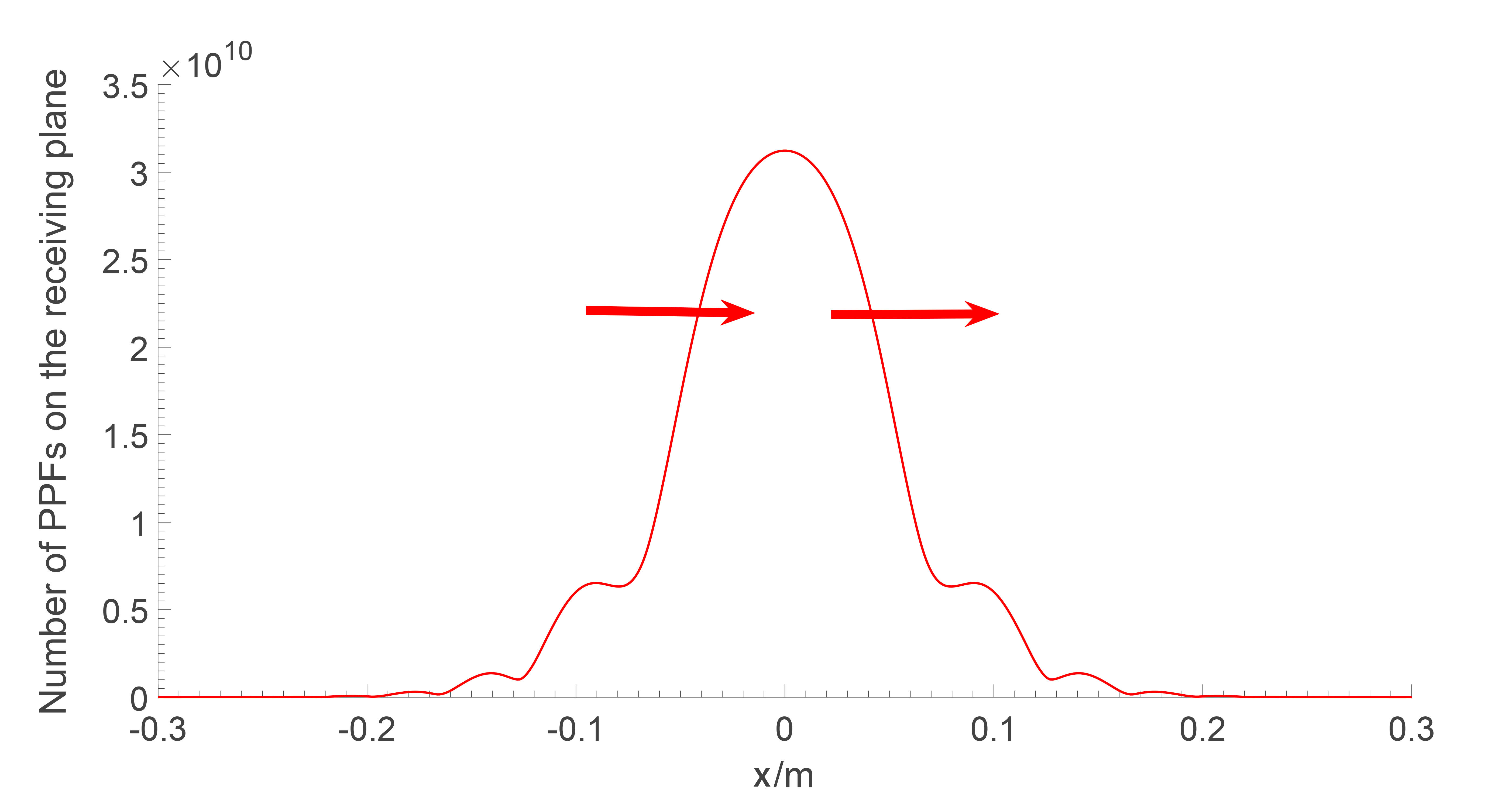}
		\caption{The transverse signal photon flow $n_{\hat x-\otimes}^{(1)}$ caused by the pure $\otimes$-type polarization of the HFGW expected by the braneworld models~\cite{Clarkson:2006pq,Seahra:2004fg}. It is shown that $n_{\hat x- \otimes}^{(1)}$ [see Eq.~(\ref{23})] is an even function of the coordinate $ x$, so $n_{\hat x-\otimes}^{(1)}$ has the same propagating direction in the regions of $ x>0$ and $ x<0$.}  \label{F7}			
	\end{figure}

In Refs.~\cite{Li:2017jcz,Li:2008qr}, we have discussed the conditions for displaying the PPFs in the BPF fluctuation, but the numerical estimation of the transverse PPFs at the different receiving surfaces were not presented. In this section, we shall give the conditions for displaying the PPFs of several typical HFGWs and the corresponding best receiving surface positions. It is found that if we consider the opposite propagating directions (e.g. see Figs.~\ref{F6} and \ref{F7}) between the PPFs and the BPFs at some receiving surfaces, the conditions for displaying the PPFs can be further relaxed.
	
In our 3DSR system, the PPFs are always accompanied by the BPF caused by the GB and other noise photons. In order to probe the PPFs, it must satisfy the following condition for displaying the PPFs in the BPF fluctuation:
	\begin{align}
		N^{(1)}_{\hat x}\Delta t  \geq \sqrt{N_{\hat x}^{(0)} \Delta t},
	\end{align}	
or
	\begin{align}\label{64xx}
		\Delta t \geqslant \frac{N_{\hat x}^{(0)}}{\left(N_{\hat x}^{(1)}\right)^2}=\Delta t_{\text {min }},
	\end{align}	
	where $\Delta t_{\text {min}}$ is the requisite minimal accumulation time of the signals. Here,
	\begin{align}
		N^{(1)}_{\hat x}=\int_{\Delta S} n^{(1)}_{\hat x} dy dz
	\end{align}	
	and
	\begin{align}
		N^{(0)}_{\hat x}=\int_{\Delta S} n^{(0)}_{\hat x} dy dz
	\end{align}
represent the total transverse signal photon flow and the total transverse noise photon flow passing through the receiving surface $\Delta S$, respectively. It should be pointed out that although these noise photon flows include the BPF caused by the GB and other various noise photon flows (such as shot noise, Johnson noise, quantization noise, thermal noise, preamplifier noise, diffraction noise, etc.), the latter ones are much less than the BPF caused by the GB if the operating temperature $T \leqslant 1\,$ K~\cite{Woods2011}. Thus, we can mainly focus on the BPF fluctuation caused by the GB. In other words, once the PPFs can be displayed in the BPF fluctuation caused by the GB, the influence of all other noise photon flow fluctuation would be negligible.
	
\begin{figure}[H]
		\centering
		\renewcommand{\figurename}{Fig.}
		\includegraphics[width=16cm]{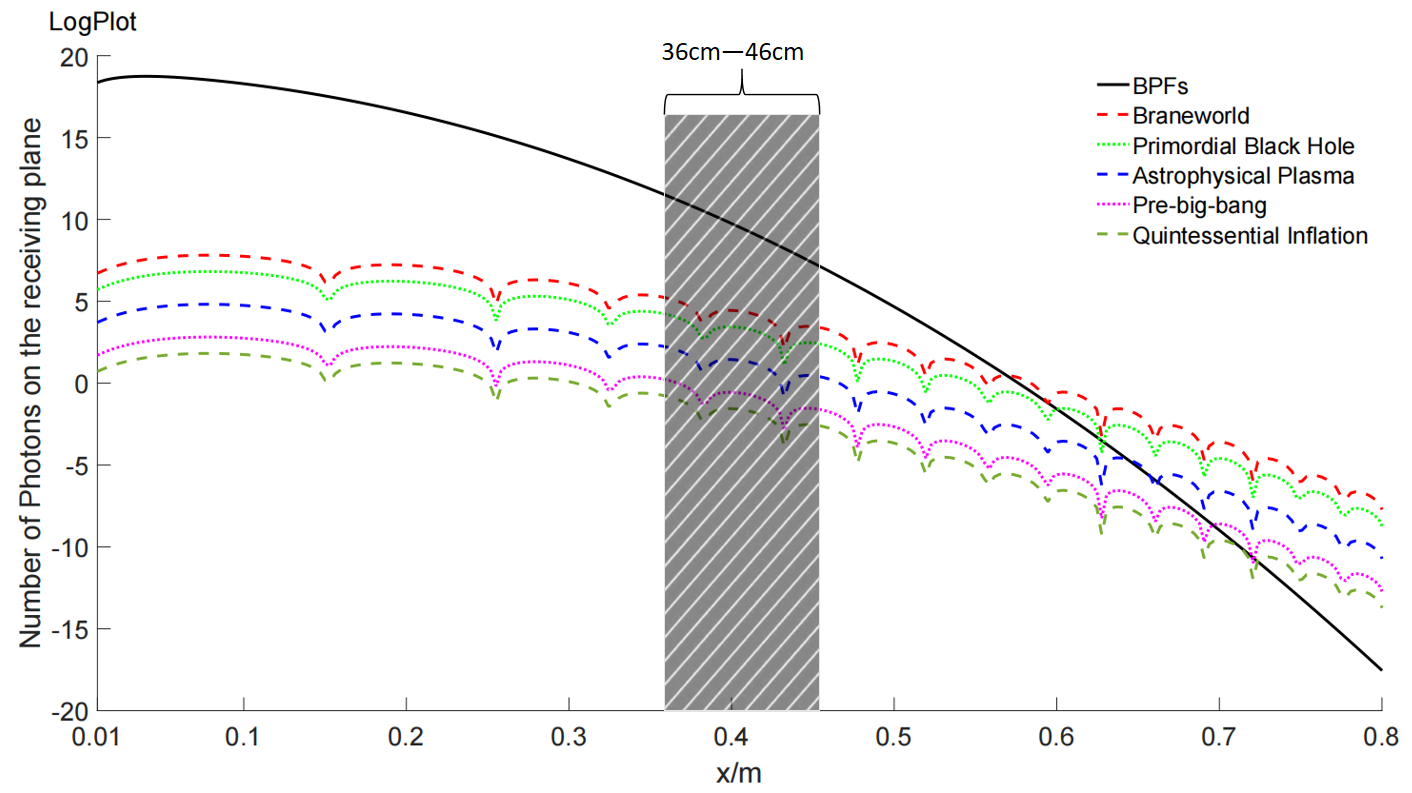}
		\caption{The total transverse signal photon flows $N^{(1)}_{\hat x}$ at the revering surface $\Delta S \sim 3 \times 10^{-2} \mathrm{~m}^2$ generated by several typical HFGWs and the total transverse background noise photon flow $N^{(0)}_{\hat x}$ caused by the GB. The figure shows that when the receiving surface  $\Delta S \sim 3 \times 10^{-2} \mathrm{~m}^2$ is within the range of $\sim36$ cm to 46 cm (about 6 to 8 times the spot radius of the GB) from the symmetry plane ($yz$-plane) of the GB, the requisite minimal accumulation time of the signals for the PPFs generated by the HFGWs from the brane oscillation, the primordial black holes and the astrophysical plasma oscillation can be limited to $10^6$\,s or less. Note that the receiving surface at $x\sim 60$ cm is significant (about 10 times the spot radius of the GB), because for this distance the PPFs generated by the HFGWs from the brane oscillation, the primordial black holes and the astrophysical plasma oscillation, etc., are the same order of magnitude as the BPF, which is satisfactory for detecting these HFGWs. For the PPFs produced by the primordial HFGWs in the Pre-big bang and the quintessential inflation, displaying them is still facing great challenges. However, since the PPFs and the BPF propagate along the opposite directions (e.g., see the regions of $x<0$ in Figs.~\ref{F6} and~\ref{F7} ), the displaying conditions for them would be relaxed greatly. Here, the ratio of the central brightness of the GB in the 3DSR system to its spot radius is a typical parameter of the ordinary GB~\cite{Yariv1989}. Therefore, there is no strict requirement for the geometric purity of the GB.}	\label{F8}		
	\end{figure}

From Figs.~\ref{F6} and \ref{F7}, one can find that the propagating directions between $n_{\hat x-\otimes}^{(1)}$ and $n_{\hat x}^{(0)}$ are opposite in the regions of $x<0$. {At this point, highly oriented receivers, such as fractal films~\cite{Wen2112,zhou1112,Li2009}, can be employed to discern variations in the propagation direction. This means that the receiver can be oriented towards a specific direction to receive the signal photon flow, thus facilitating the reception of the signal photon flow. Consequently, it becomes feasible to reduce the intensity of the noise photon flow passing through the receiving surface $\Delta S$ while maintaining the intensity of the signal photon flow. Therefore, the accumulation time of the displayed signal [see Eq.~(\ref{64xx})] can be reduced, leading to an enhancement in sensitivity.} Therefore, considering the different strength distributions, the decay rates, and the wave impedances between $n_{\hat x-\otimes}^{(1)}$ and $n_{\hat x}^{(0)}$, it is quite possible to distinguish and probe the signal photon flows produced by the HFGWs expected by the braneworld models~\cite{Clarkson:2006pq,Seahra:2004fg}, the primordial black holes~\cite{Aggarwal:2020olq}, and the interaction between the astrophysical plasma and the intense EM radiation~\cite{Servin:2003cf} due to their larger amplitudes (or higher spectral densities) and the spectral characteristics. At present, distinguishing and detecting the signal photon flows generated by the primordial (relic) HFGWs such as the Pre-big bang~\cite{Gasperini:2002bn,Veneziano}, the quintessential inflation~\cite{Giovannini:1999bh}, etc., are still facing great challenges, but they are not impossible (see Fig.~\ref{F8}).

\begin{sidewaystable}[h]
	\caption{Displaying conditions for the PPFs in the 3DSR system (in the laboratory frame system) to the HFGWs from some typical cosmological models and high-energy astrophysical processes, where $N_{\hat x}^{(1)}$ is the total transversal signal photon flow at the receiving surface ($\Delta S\sim 3\times 10^{-2}$ m$^2$), $N_{\hat x}^{(0)}$ is the allowable upper limit of the total transverse noise photon flow at the receiving surface, and $\Delta t_{\text{min}}$ is the requisite minimal accumulation time of the signals.}~\\  
	\centering 
	\begin{tabular}{l c c rrrr} 
		\hline\hline 
		Theories and  & Braneworld& Primordial & Short-term & Interaction of  & Pre-big bang & Quintessential
		\\[-1ex]
		
		\raisebox{0.5ex}{models $\&$ main} & \raisebox{0.5ex}{~\cite{Clarkson:2006pq,Seahra:2004fg}}&\raisebox{0.5ex}{black holes~\cite{Aggarwal:2020olq,Franciolini:2022htd}}&\raisebox{0.5ex}{anisotropic }&\raisebox{0.5ex}{astrophysical plasma }&\raisebox{0.5ex}{\cite{Gasperini:2002bn,Veneziano}~~~~~~~} &\raisebox{0.5ex}{inflation or upper}\\[-1ex]
		
		\raisebox{0.5ex}{ issues and } & \raisebox{0.5ex}{}&\raisebox{0.5ex}{}&\raisebox{0.5ex}{inflation~\cite{Ito:2016aai}}&\raisebox{0.5ex}{with EM radiation}&\raisebox{0.5ex}{} &\raisebox{0.5ex}{ limit of ordinary}	\\
		[-1ex]
		
		\raisebox{0.5ex}{  parameters} & \raisebox{0.5ex}{}&\raisebox{0.5ex}{}&\raisebox{0.5ex}{}&\raisebox{0.5ex}{\cite{Servin:2003cf}~~~~~~~~~~~~}&\raisebox{0.5ex}{} &\raisebox{0.5ex}{ inflation~\cite{Tong:2009vk,Giovannini:1999bh}}	\\

		\hline
		Basic  & Extra-dimensions  & Very early   & Very early &  The intraction &  The start point & The essence and  \\[-1ex]
		
		\raisebox{0.5ex}{scientific} &
		\raisebox{0.5ex}{of space}&\raisebox{0.5ex}{universe and}&\raisebox{0.5ex}{universe and}&\raisebox{0.5ex}{mechanism between}&\raisebox{0.5ex}{of time or the} &\raisebox{0.5ex}{candidates of}
		\\[-1ex]
		
		\raisebox{0.5ex}{issues} & \raisebox{0.5ex}{and multiple}&\raisebox{0.5ex}{ black hole}&\raisebox{0.5ex}{inflationary}&\raisebox{0.5ex}{ astrophysical}&\raisebox{0.5ex}{information from} &\raisebox{0.5ex}{dark energy  }
		\\[-1ex]
		
		\raisebox{0.5ex}{involved} & \raisebox{0.5ex}{brane universes}&\raisebox{0.5ex}{ physics}&\raisebox{0.5ex}{epoch}&\raisebox{0.5ex}{plasma and intnese}&\raisebox{0.5ex}{the Pre-big bang } &\raisebox{0.5ex}{ and inflation}
		\\[-1ex]
		
		\raisebox{0.5ex}{} & \raisebox{0.5ex}{}&\raisebox{0.5ex}{}&\raisebox{0.5ex}{}&\raisebox{0.5ex}{ EM radiation}&\raisebox{0.5ex}{} &\raisebox{0.5ex}{}
		\\[-1ex]
		
		\raisebox{0.5ex}{} & \raisebox{0.5ex}{}&\raisebox{0.5ex}{}&\raisebox{0.5ex}{}&\raisebox{0.5ex}{}&\raisebox{0.5ex}{} &\raisebox{0.5ex}{}	\\

		Amplitude (A) & $\sim 10^{-21}$ to $10^{-23}$ &$10^{-24}$ to $\sim 10^{-25}$   & $\sim 10^{-26}$   & $\sim 10^{-27}$  & $\sim 10^{-29}$  & $\sim 10^{-30}$ or less   \\[-1ex]
		\raisebox{0.5ex}{(dimensionless)} & \raisebox{0.5ex}{}&\raisebox{0.5ex}{upper limit}&\raisebox{0.5ex}{}&\raisebox{0.5ex}{}&\raisebox{0.5ex}{} &\raisebox{0.5ex}{}\\
		\\[-1ex]
		
		\raisebox{0.5ex}{} & \raisebox{0.5ex}{}&\raisebox{0.5ex}{}&\raisebox{0.5ex}{}&\raisebox{0.5ex}{}&\raisebox{0.5ex}{} &\raisebox{0.5ex}{}
		\\
		
		Resonance & $\sim 10^9~~\text{to}~~10^{12}$ & $\sim 10^9~~\text{to}~~10^{10}$  & $\sim 10^8~~\text{to}~~10^{9}$  & $\sim 10^{11}~~\text{to}~~10^{12}$ & $\sim 10^9~~\text{to}~~10^{10}$  & $\sim 10^9~~\text{to}~~10^{10}$  \\[-1ex]
		\raisebox{0.5ex}{Frequency (Hz)} & \raisebox{0.5ex}{}&\raisebox{0.5ex}{}&\raisebox{0.5ex}{}&\raisebox{0.5ex}{}&\raisebox{0.5ex}{} &\raisebox{0.5ex}{}\\
		\\[-1ex]
		
		\raisebox{0.5ex}{} & \raisebox{0.5ex}{}&\raisebox{0.5ex}{}&\raisebox{0.5ex}{}&\raisebox{0.5ex}{}&\raisebox{0.5ex}{} &\raisebox{0.5ex}{}
		\\

		& Discrete  & Continuous   & Stochastic & Stochastic & Stochastic  & Stochastic  \\[-1ex]
		\raisebox{0.5ex}{Spectral} & \raisebox{0.5ex}{spectrum}&\raisebox{0.5ex}{spectrum}&\raisebox{0.5ex}{background}&\raisebox{0.5ex}{background}&\raisebox{0.5ex}{background} &\raisebox{0.5ex}{background}
		\\[-1ex]
		
		\raisebox{0.5ex}{characteristic} & \raisebox{0.5ex}{(coherent}&\raisebox{0.5ex}{(coherent}&\raisebox{0.5ex}{}&\raisebox{0.5ex}{}&\raisebox{0.5ex}{} &\raisebox{0.5ex}{}
		\\[-1ex]
		
		\raisebox{0.5ex}{} & \raisebox{0.5ex}{HFGWs)}&\raisebox{0.5ex}{HFGWs)}&\raisebox{0.5ex}{}&\raisebox{0.5ex}{}&\raisebox{0.5ex}{} &\raisebox{0.5ex}{}
		\\[-1ex]
		
		\raisebox{0.5ex}{} & \raisebox{0.5ex}{}&\raisebox{0.5ex}{}&\raisebox{0.5ex}{}&\raisebox{0.5ex}{}&\raisebox{0.5ex}{} &\raisebox{0.5ex}{}	\\	
		
		$N_{\hat x}^{(1)}$ (s$^{-1}$) & $\sim 10^{9}$ to $\sim 10^{11}$ & $\sim 10^{8}$ & $\sim 10^{6}$ & $\sim 10^{5}$ & $\sim 10^{3}$ & $\sim 10^{2}$ \\[-1ex]
		
		\raisebox{0.5ex}{} & \raisebox{0.5ex}{}&\raisebox{0.5ex}{}&\raisebox{0.5ex}{}&\raisebox{0.5ex}{}&\raisebox{0.5ex}{} &\raisebox{0.5ex}{}\\

		$\Delta t_{\text{min}}$(s) & $10^{4}$ or less   &  $\sim 10^6$ or less &  $\sim 10^5$ &  $\sim 10^5$ &  $\sim 10^6$ &  $\sim 10^6$ or larger \\[-1ex]

		\raisebox{0.5ex}{} & \raisebox{0.5ex}{}&\raisebox{0.5ex}{}&\raisebox{0.5ex}{}&\raisebox{0.5ex}{}&\raisebox{0.5ex}{} &\raisebox{0.5ex}{}\\

		$N_{\hat x}^{(0)}$ (s$^{-1}$)& $\sim 10^{22}$ & $\sim 10^{22}$ & $\sim 10^{18}$ & $\sim 10^{16}$ & $\sim 10^{12}$ & $\sim 10^{10}$ \\[-1ex]
		
		\raisebox{0.5ex}{} & \raisebox{0.5ex}{}&\raisebox{0.5ex}{}&\raisebox{0.5ex}{}&\raisebox{0.5ex}{}&\raisebox{0.5ex}{} &\raisebox{0.5ex}{}\\
		\hline
	\end{tabular}
	\label{tab1}
\end{sidewaystable}

{Certainly, even if the 3DSR system attains the aforementioned ideal state, the presence of reflection noise resulting from the reflection of photon flow remains a major challenge. To suppress these noise, several highly beneficial and innovative ideas and suggestions have been proposed. For example, placing effective absorbing microwave ``blackbody materials'' on the cavity wall, increasing the transverse distance between the reflecting surface and the Gaussian beam, applying highly directional selective fractal films (an equivalent microwave lens), adopting high-quality GB, using deep learning and neural network methods for signal analysis, etc~\cite{Woods2011,ShahzadAnwar,Wen2112,zhou1112,Li2009,XiuLin,Ringwald:2020ist,Wang:2019ybv}.}

Table~\ref{tab1} lists the conditions for displaying the total transverse signal photon flow $N^{(1)}_{\hat x}$ at the receiving surface $\Delta S\left(\sim 3 \times 10^{-2} \mathrm{~m}^{2}\right)$ for various situations in our 3DSR system.  $N_{\hat x}^{(0)}$ is the allowable upper limit of the total noise photon flow at the receiving surface $\Delta S$, and $\Delta t_{\text{min}}$ (see Figs.~\ref{F6} and \ref{F7}) is the requisite minimal accumulation time of the signals [see Eq.~(\ref{64xx})].

In our 3DSR system, $\hat B^{(0)}=10 $ T (the background static magnetic field), the power of the GB is $\sim 10 \mathrm{~W}$, and the  operating temperature is less than $1\,$K. Therefore, the maximum of $N_{\hat x}^{(0)}$ at the receiving surface $\Delta S\left(\sim 3 \times 10^{-2} \mathrm{~m}^{2}\right)$ is about $\sim 10^{18} \mathrm{~s}^{-1}$. Thus, even if the peak values of the noise photon flow and the PPFs appear at the same receiving surface (see Table~\ref{tab1}), $\Delta t_{\text{min}}$ can be limited to $\sim 10^{6} \mathrm{~s}$ or less for the HFGWs expected by the braneworld models~\cite{Clarkson:2006pq,Seahra:2004fg}, the astrophysical plasma oscillation~\cite{Servin:2003cf}, and the primordial black hole theories~\cite{Aggarwal:2020olq}. However, for most cases in our 3DSR system, the peak positions of the two kinds of photon flows do not appear at the same receiving surface. Moreover, if we consider the very different physical behaviors (such as the strength distribution, the propagating direction, the decay rate and the wave impedance) between $n_{\hat x}^{(0)}$ and $n_{\hat x}^{(1)}$ in some special local regions, the displaying conditions for these HFGWs can be further relaxed.

	\section{Concluding remarks}

In this paper, we study the EM resonance response to the HFGWs having additional polarization states in the laboratory frame system, and show that the PPFs generated by the tensor and non-tensor polarization states of HFGWs have separability and detectability. Our conclusions can boil down to the following points.
	
(1) In general, the intensities and propagating directions of the signal photon flows in the EM resonance response depend on the relative position of the 3DSR system and the standard GW frame system. When the propagating direction of HFGWs and the direction of the symmetrical axis of the 3DSR system are the same, the transverse signal photon flows have a maximum value. In this case, the difference between the EM perturbations in the standard GW frame system and the ones in the laboratory frame system will be the first-order EM perturbations $\sim h$ [see Eqs.~(\ref{newb22}) and (\ref{newb23})].
	
(2) Under the above EM resonance response, the coordinate rotation (the rotation of the azimuth $\phi$) only causes the conversion between the $\otimes$-type and $\oplus$-type polarizations (the tensor polarizations), and the conversion between the $x$-type and $y$-type polarizations (the vector polarizations). Therefore, there is no conversion between the tensor polarizations and the non-tensor polarizations. Also, there is no conversion between the vector polarizations and the scalar polarizations.
	
(3) In the standard GW frame system (i.e., $\theta=\phi=0$), the pure $\otimes$-type polarization, the pure $x$-type polarization and the pure $y$-type polarization can be displayed by the PPFs $n^{(1)}_{\hat x-\otimes}$, $n^{(1)}_{\hat x-x}$ and $n^{(1)}_{\hat x-y}$, respectively. The combination state of the $\oplus$-type and $l$-type polarizations, and the combination state of the $\otimes$-type, $b$-type and $l$-type polarizations can be displayed by the PPFs $n_{\hat y-\oplus,l}^{(1)}$ and $n_{\hat x-\otimes, b, l}^{(1)}$, respectively. Moreover, the coupling between the longitudinal static magnetic field $\hat{B}_{z}^{(0)}$ and the GB can display the pure additional polarizations (the $x$-type and $y$-type polarizations). Therefore, this is a useful way to test the gravity theory beyond GR.
	
(4) In the laboratory frame system $(\theta=0, \phi=\pi / 4$ and $\pi / 2)$, the pure $\oplus$-type polarization, the pure $x$-type polarization and the pure $y$-type polarization can be displayed by the PPFs $n_{\hat x-\oplus}^{(1)}$, $n_{\hat x-x}^{(1)}$ and $n_{\hat x-y}^{(1)}$, respectively. The combination state of the $\oplus$-type and $l$-type polarizations, and the combination state of the $\otimes$-type and $l$-type polarizations can be displayed by the PPFs $n_{\hat y-\oplus,l}^{(1)}$ and $n_{\hat y-\otimes,l}^{(1)}$, respectively. According to the above analysis, the pure $l$-type polarization can be displayed by the PPF $n_{\hat y-l}^{(1)}$.
	
If we consider the different physical behaviors between the transverse PPFs and the BPFs in some special local regions, such as the intensity distribution, the decay rate, the wave impedance, especially, the opposite propagating direction (see Figs.~\ref{F6} and \ref{F7}), the displaying conditions for the PPFs would be greatly relaxed. Our numerical estimations show that at about 6 to 8 times the spot radius of the GB ($\sim36$ cm to 46 cm) from the symmetry plane ($yz$-plane), the PPFs have best displaying conditions, where the minimal accumulation time of the signals would be reduced to $10^{6} \mathrm{~s}$ or less for some high-energy HFGWs. Moreover, all six polarization states of HFGWs would have separability and detectability.
	
(5) The numerical estimations show that the PPFs generated by the HFGWs from the braneworld models, the primordial black holes, the high energy astrophysical plasma, etc., have best detectability due to the very high frequency, the large amplitudes (or the high spectral densities) and the coherent property. The detection of the primordial HFGWs expected by some inflationary models faces great challenges due to the small amplitudes (or the low spectral densities), but it is not impossible.
	
Finally, it should be pointed out that there are still some important and interesting issues which are worthy of future study. (1) Since the positions of some possible coherent HFGW sources are uncertain, it will lead to the uncertainty of the intensities of the PPFs in the 3DSR system. Therefore, it is an important project to seek the general relationship between the orientations of the HFGW sources and the intensities of the PPFs, including the 24-hour periodic change of the intensities of the PPFs due to the rotation of the Earth. In this case, a possible solution is to use three or more 3DSR systems in different locations to carry out relevant coincidence experiments. (2) In fact, what is shown in Figs.~\ref{F6} to \ref{F8} is only the intensities of the signal photon flow and the background noise photon flow. Clearly, if the characteristics of the spectrum of HFGWs can be included in the figures, then the effects will be further improved (e.g., the discrete spectrum of the HFGWs from the braneworld models, the continuous spectrum of the HFGWs from the primordial black holes, etc.). (3) The law of conservation of angular momentum for the EM resonance response to HFGWs in the GR framework system, strongly implies the existence of the massless spin-2 gravitons. These issues will be discussed and studied in detail elsewhere.

About 50 years ago (1973), C. W. Misner, Kip. S. Thorne and J. A. Wheeler once pointed out that ``But whether they do or not, gravitational wave astronomy has begun and seems to have a bright future"~\cite{Misner:1973prb}. Today, the LIGO Scientific Collaboration and the Virgo Collaboration have successfully fulfilled this desire. If people hope to observe more information on cosmology and high-energy astrophysics, and find a way to realize the quantization of gravity from an observational perspective, it is necessary for us to focus on the effect of HFGWs (i.e., high-energy gravitons). When there is significant breakthrough in the observation of HFGWs in the future, it is not only expected to be an important window for unifying quantum theory and classical gravity theory, but also lead us to a more fundamental physical background than spacetime.
	
\section*{Acknowledgments} \hspace{5mm}
This project is supported by the National Natural Science Foundation of China (Grant No.11375279, No.11605015, No.12047564 and No.11873001), the Fundamental Research Funds for the Central Universities (Grand No. 2021CDJZYJH-003), and the Postdoctoral Science Foundation of Chongqing (Grant No. cstc2021jcyj-bsh0124). We thank helpful discussions with Prof. Robert M. L. Baker, Prof. R. C. Woods, Prof. C. Corda, Dr. Gary V. Stephenson, Dr. Andrew Beckwith and Dr. Eric W. Davis. It is with great sadness that the passing of Professor Bob Baker in January 2023. He has made outstanding contributions to the cooperative research between the China-US high-frequency gravitational wave scientific teams.

\end{document}